\pdfoutput=1
\newcommand*{\ATLASLATEXPATH}{latex/}
\documentclass[cernpreprint,texlive=2016,txfonts,UKenglish]{latex/atlasdoc}
\usepackage[backend=bibtex]{\ATLASLATEXPATH atlaspackage}
\usepackage{\ATLASLATEXPATH atlasbiblatex}
\usepackage{\ATLASLATEXPATH atlascontribute}
\usepackage{\ATLASLATEXPATH atlasphysics}
\usepackage{\ATLASLATEXPATH atlasheavyion}
\usepackage{\ATLASLATEXPATH atlasparticle}
\usepackage{\ATLASLATEXPATH atlasprocess}
\graphicspath{{logos/}{figures/}}

\usepackage{atlas-document-defs}
\usepackage{multirow}

\AtlasTitle{Measurement of quarkonium production in proton--lead and proton--proton collisions at $5.02~\mathrm{TeV}$ with the ATLAS detector}
\author{The ATLAS Collaboration}
\PreprintIdNumber{CERN-EP-2017-197}
\AtlasDate{\today}
\AtlasJournal{Eur. Phys. J. C}
\AtlasJournalRef{Eur. Phys. J. C 78 (2018) 171}
\AtlasDOI{10.1140/epjc/s10052-018-5624-4}
 
\AtlasAbstract{
  The modification of the production of $J/\psi$, $\psi(2\mathrm{S})$, and $\mit{\Upsilon}(n\mathrm{S})$ ($n = 1, 2, 3$) 
  in $p$+Pb collisions with respect to their production in $pp$ collisions has been studied. 
  The $p$+Pb and $pp$ datasets used in this paper correspond to integrated luminosities of $28~\mathrm{nb}^{-1}$ and $25~\mathrm{pb}^{-1}$ respectively, 
  collected in 2013 and 2015 by the ATLAS detector at the LHC, both at a centre-of-mass energy per nucleon pair of 5.02 TeV. 
  The quarkonium states are reconstructed in the dimuon decay channel. 
  The yields of $J/\psi$ and $\psi(\mathrm{2S})$ are separated into prompt and non-prompt sources. 
  The measured quarkonium differential cross sections are presented as a function of rapidity and transverse momentum, 
  as is the nuclear modification factor, $R_{p\mathrm{Pb}}$ for $J/\psi$ and $\mit{\Upsilon}(n\mathrm{S})$. 
  No significant modification of the $J/\psi$ production is observed while $\mit{\Upsilon}(n\mathrm{S})$ production is found 
  to be suppressed at low transverse momentum in $p$+Pb collisions relative to $pp$ collisions. 
  The production of excited charmonium and bottomonium states is found to be suppressed relative to that of the ground states in central $p$+Pb collisions.
}

\hypersetup{pdftitle={ATLAS document},pdfauthor={The ATLAS Collaboration}}

\begin{document}
\maketitle

\section{Introduction}
\label{sec:intro}

The study of heavy quarkonium bound states ($c\bar{c}$ and $b\bar{b}$) in ultra-relativistic heavy-ion collisions~\cite{Lansberg:2008zm,Brambilla:2010cs}
has been a subject of intense theoretical and experimental efforts since it was initially proposed by Matsui and Satz~\cite{Matsui1986416} as a probe to study a deconfined
quark--gluon plasma (QGP) created in nucleus--nucleus (A+A) collisions. 
In order to understand quarkonium yields in A+A collisions 
it is necessary to disentangle effects due to interaction between quarkonium and the QGP medium from those that can be ascribed to cold nuclear matter (CNM). 
In proton (deuteron)--nucleus collisions,  $p$($d$)+A, the formation of a large region of deconfined and hot QGP matter was not expected to occur.
Therefore, the observed suppression of quarkonium yields in these systems with respect to $pp$ collisions~\cite{Alessandro:2003pc,Adler:2005ph,PhysRevC.77.024912,PhysRevLett.107.142301} 
has traditionally been attributed to CNM effects. 

Among the CNM effects,
three primary initial-state effects are: 
modifications of the nuclear parton distribution functions~\cite{PhysRevD.11.3105,NPA.696.669,PhysRevC.71.054902,PhysRevC.87.014908}, 
parton saturation effects in the incident nucleus~\cite{FUJII20131},
and parton energy loss through interactions with the nuclear medium~\cite{PhysRevLett.68.1834,JHEP.03.122}.
On the other hand, the absorption of the heavy quark--antiquark pair through interactions with the co-moving nuclear medium~\cite{Capella1988354,PLB.366.316,Vogt1999197,Elena2014}
is considered to be a final-state effect.
In proton--lead (\pPb) collisions, the modification of quarkonium production with respect to that in $pp$ collisions may be quantified by the nuclear modification factor, \rpb,
which is defined as the ratio of the quarkonium production cross section in \pPb\  collisions 
to the cross section measured in $pp$ collisions at the same centre-of-mass energy,
scaled by the number of nucleons in the lead nucleus:

\begin{equation*}
       \rpb = \frac{1}{208}\frac{\sigma_{\pPb}^{\mathcal{O}(n\text{S})}  }{\sigma_{pp}^{\mathcal{O}(n\text{S})} },
\end{equation*}
where $\mathcal{O}(n\text{S})$ represents one of five measured quarkonium states, $\Jpsi$, $\psip$, $\Upsi$, $\Upsii$ and $\Upsiii$. 
Several measurements of CNM effects in quarkonium production were performed with \pPb\ data collected in 2013 at the LHC
at a centre-of-mass energy per nucleon pair \energy.
Measurements of the $\Jpsi$ nuclear modification factor and forward ($p$ beam direction) to backward ($\mathrm{Pb}$ beam direction) cross-section ratio 
by the ALICE~\cite{Abelev:2013yxa,ALICE_pPbJpsi2} and LHCb~\cite{LHCb_pPbJpsi} experiments show strong suppression at large rapidity and low transverse momentum.
However, no strong modification of $\Jpsi$ production is observed at small rapidities and high transverse momentum 
by the ATLAS~\cite{HION-2013-07} or CMS~\cite{Sirunyan2017} experiments
indicating that the CNM effects have strong rapidity and/or transverse momentum dependence.
The CNM effects in excited quarkonium states with respect to the ground state can be quantified by the double ratio, \rhopb,
defined as:

\begin{equation*}
       \rhopb = \frac{ \rpb(\mathcal{O}(n\text{S})) }{ \rpb(\mathcal{O(\text{1S})}) } = \frac{ \sigma_{\pPb}^{\mathcal{O}(n\text{S})} }{\sigma_{\pPb}^{\mathcal{O(\text{1S})}}} / \frac{ \sigma_{pp}^{\mathcal{O}(n\text{S})} }{\sigma_{pp}^{\mathcal{O(\text{1S})}}},
\label{eq:DR_def}
\end{equation*}
where $n = 2$ for charmonium and $n = 2~\text{or}~3$ for bottomonium.
In the double ratio, most sources of detector systematic uncertainty cancel out, and measurements of this quantity by different experiments can easily be compared.
The initial-state effects are expected to be largely cancelled out in double ratio due to the same modifications affecting partons before the formation of the quarkonium state,
so measuring the relative suppression of different quarkonium states should help in understanding the properties of the final-state effects separately from the initial ones.
The PHENIX experiment at RHIC has presented measurements of $\psip$ suppression at mid-rapidity for $d$+Au interactions at $\sqn = 200~\GeV$,
showing that the charmonium double ratio is smaller than unity,
and decreases from peripheral to central collisions~\cite{Adare:2013ezl}.
At the LHC, inclusive $\Jpsi$~\cite{Abelev:2013yxa} and $\psip$~\cite{Abelev:2014zpa} production has been measured by the ALICE experiment in \pPb\ collisions at $\energy$ at forward rapidity.
Those measurements show a significantly larger suppression of the $\psip$ compared to that measured for $\Jpsi$.

The $\Upsins$ ($n=2,3$) to $\Upsi$ double ratios are both found to be less than unity by the CMS experiment in \pPb\ collisions at $\energy$~\cite{CMS-HIN-13-003}.
A double ratio which is smaller than unity suggests the presence of final-state interactions that affect the excited states more strongly than the ground state,
since initial-state effects are expected to cancel.
The CMS \pPb\ results indicate that the CNM effect partially contributes to the strong relative suppression 
found in previous CMS measurements~\cite{CMS-HIN-11-007, CMS-HIN-11-011, CMS-HIN-15-001} of \PbPb collisions at $\sqn = 2.76~\TeV$.

In this paper, four classes of experimental measurements are presented.
The first class of measurements is differential production cross sections of $\Jpsi$, $\psip$, and $\Upsins$ ($n = 1, 2, 3$) 
in $pp$ collisions at $\ppenergy$ and \pPb collisions at $\energy$.
The second is the centre-of-mass rapidity dependence and transverse momentum dependence of $\Jpsi$ and $\Upsi$ nuclear modification factors, $\rpb$.
The third is the evolution of the quarkonium yields with \pPb collision centrality~\cite{HION-2012-15}
studied using ratios of the yields of quarkonia to that of $Z$ bosons and 
the correlation between quarkonium yields and event activity, where both are normalised by their average values over all events.
The fourth is the charmonium and bottomonium double ratios, \rhopb , presented as a function of centre-of-mass rapidity and centrality.

\section{ATLAS detector}
\label{sec:detector}
\newcommand{\AtlasCoordFootnote}{%
ATLAS uses a right-handed coordinate system with its origin at the nominal interaction point (IP) in the centre of the detector and the $z$-axis along the beam pipe. The $x$-axis points from the IP to the centre of the LHC ring, and the $y$-axis points upward. Cylindrical coordinates $(r,\phi)$ are used in the transverse plane, $\phi$ being the azimuthal angle around the beam pipe.
The pseudorapidity is defined in terms of the polar angle $\theta$ as $\eta = -\ln \tan(\theta/2)$  
and the transverse momentum \pt is defined
as $\pt = p\sin\theta$. 
Angular distance is measured in units of $\Delta R \equiv \sqrt{(\Delta\eta)^{2} + (\Delta\phi)^{2}}$.
}

The ATLAS experiment~\cite{PERF-2007-01} at the LHC is a multi-purpose detector
with a forward-backward symmetric cylindrical geometry and a nearly $4\pi$ coverage in
solid angle.\footnote{\AtlasCoordFootnote}
It consists of an inner tracking detector (ID) surrounded by a thin superconducting solenoid
providing a \SI{2}{\tesla} axial magnetic field, electromagnetic and hadron calorimeters, and a muon spectrometer (MS).
The ID covers the pseudorapidity range $|\eta| < 2.5$.
It consists of silicon pixel, 
silicon micro-strip, and gaseous transition radiation tracking detectors.
A new innermost insertable B-layer~\cite{ATLAS-TDR-19} installed during the first LHC long shutdown (2013 to 2015) 
has been operating as a part of the silicon pixel detector since 2015.
The calorimeter system covers the pseudorapidity range $|\eta| < 4.9$. Within the region $|\eta| < 3.2$, electromagnetic
calorimetry is provided by barrel and endcap high-granularity lead/liquid-argon (LAr) electromagnetic
calorimeters, with an additional thin LAr presampler covering $|\eta| < 1.8$ to correct for energy
loss in material upstream of the calorimeters. Hadronic calorimetry is provided by a steel/scintillator tile
calorimeter, segmented into three barrel structures within $|\eta| < 1.7$, and two copper/LAr hadronic
endcap calorimeters. The solid angle coverage is completed with forward copper/LAr and tungsten/LAr
calorimeter (FCal) modules optimised for electromagnetic and hadronic measurements respectively.
The MS comprises separate trigger and high-precision tracking chambers measuring the deflection of muons 
in a magnetic field generated by superconducting air-core toroids.
Monitored drift tubes and cathode strip chambers
are designed to provide precise position measurements in the bending plane in the range $|\eta| < 2.7$.
Resistive plate chambers (RPCs) and thin gap chambers (TGCs) with a coarse position resolution
but a fast response time are used primarily to trigger on muons in the ranges
$|\eta| < 1.05$ and $1.05 < |\eta| < 2.4$ respectively.

The ATLAS trigger system~\cite{TRIG-2012-03, ATLAS_TrigPerf2015}
is separated into two
levels: the hardware-based level-1 (L1) trigger and
the software-based high level trigger (HLT), which reduce the 
proton--proton/lead collision rate to several-hundred Hz of events of interest for
data recording to mass storage.
The L1 muon trigger requires coincidences between hits on different RPC or TGC planes,
which are used as a seed for the HLT algorithms.
The HLT uses dedicated algorithms
to incorporate information from both the MS and the ID,
achieving position and momentum resolution close to that
provided by the offline muon reconstruction, as shown in Ref.~\cite{TRIG-2012-03}.
During the first LHC long shutdown 
additional RPCs were installed to cover the acceptance holes at the bottom of the MS
and additional TGC coincidence logic was implemented for the region $1.3 < |\eta| < 1.9$ to reduce backgrounds.
More details about the improvement in the trigger system during the long shutdown can be found in Ref.~\cite{ATLAS_TrigPerf2015}.

\section{Datasets and Monte Carlo samples}
This analysis includes data from \pPb collisions recorded at the LHC in 2013 and \pp collisions recorded in 2015,
both at a centre-of-mass energy of $5.02~\TeV$ per nucleon pair.
These data samples correspond to a total integrated luminosity of $28~\inb$ and $25~\ipb$ for
\pPb and $pp$ collisions respectively.

The \pPb collisions result from the interactions of a proton beam with an energy of $4~\TeV$ and a lead beam with an energy of $1.58~\TeV$ per nucleon.
The usual rapidity, $y$, in the laboratory frame is defined as $y = 0.5 \ln[(E + p_z) / (E - p_z)]$, where $E$ and $p_z$ refer to energy and longitudinal momentum respectively.
In the \pPb collision configuration, the proton--nucleon centre-of-mass rapidity, $y^*$, had a shift of $\Delta y = 0.465$ with respect to $y$ in the laboratory frame.
After $60\%$ of the data were recorded the directions of the proton and lead beams were reversed.
In this paper, all data from both periods are presented in $y^*$, using an additional convention
that the proton beam always travels in the direction of positive $y^*$.

Monte Carlo (MC) simulations~\cite{SOFT-2010-01} of \pPb\ and $pp$ collision events 
are used to study muon trigger and reconstruction efficiencies,
and quarkonium signal yields extraction.
Events were generated using \PYTHIA8~\cite{pythia8:2008} with the CTEQ61L~\cite{Pumplin:2002vw} parton distribution functions.
In each event, one of the five quarkonium states, $\Jpsi$, $\psip$, and $\Upsins$ ($n = 1, 2, 3$), 
was produced unpolarised, 
as motivated by previous measurements at the LHC energy~\cite{CMS-BPH-11-023,CMS-BPH-13-003,LHCb_Jpsi_polarization}, 
and forced to decay via the dimuon channel.
The response of the ATLAS detector was simulated using \GEANT4~\cite{Agostinelli:2002hh}.
The simulated events were reconstructed with the same algorithms used for data.

\section{Event selection}

Candidate events in \pPb\ collisions were collected with a dimuon trigger which requires one muon 
to pass the identification requirement at L1. 
In addition, the L1 muon candidate must be confirmed in the HLT as a muon with $\pt > 2~\GeV$,
and at least one more muon candidate with $\pt > 2~\GeV$ must be found in a search over the full MS system.
In $pp$ collisions the candidate events were collected with a different dimuon trigger which requires at least two L1 muon candidates with $\pt > 4~\GeV$.
Subsequently in the HLT, the two L1 candidates must be confirmed as muons from a common vertex with opposite-sign charges.

In the offline analysis, events are required to have at least one reconstructed primary vertex with at least four tracks 
and at least two muons originating from a common vertex, each with $\pt > 4~\GeV$ and matching an HLT muon candidate associated with the event trigger.
The selected muons are required to be \emph{Combined}~\cite{PERF-2014-05}
and \emph{Tight}~\cite{PERF-2015-10}  in \pPb\ and \pp collisions respectively,
where \emph{Combined} implies that a muon results from a track in the ID which can be combined with one in the MS, 
and \emph{Tight} requires a strict compatibility between the two segments. 
To ensure high-quality triggering and accurate track measurement,
each muon track is further restricted to $|\eta| < 2.4$.
Pairs of muon candidates satisfying these quality requirements, and with opposite charges, are selected as quarkonia candidate pairs.
All candidate pairs that satisfy the criteria discussed above, 
including those events with additional interactions in the same bunch crossing (known as "pile-up" events), are used for the cross-section measurements.

In order to characterise the \pPb\ collision geometry, each event is assigned to a centrality class
based on the total transverse energy measured in the FCal on the Pb-going side (backwards).
Collisions with more (fewer) participating nucleons are referred to as central (peripheral).
Following Ref.~\cite{HION-2012-15},
the centrality classes used for this analysis, in order from most central to most
peripheral, are $0$--$5\%$, $5$--$10\%$, $10$--$20\%$, $20$--$30\%$, $30$--$40\%$, $40$--$60\%$, and $60$--$90\%$.

\section{Analysis}
\label{sec:analysis}

\subsection{Cross-section determination}

The double-differential cross section multiplied by the dimuon decay branching fraction is calculated for each measurement interval as:

\begin{equation}
    \frac{\mathrm{d}^2\sigma_{\mathcal{O}(n\text{S})}}{\mathrm{d}\pt \mathrm{d}y^*}\times B(\mathcal{O}(n\text{S})\rightarrow\mu^+\mu^-) = \frac{N_{\mathcal{O}(n\text{S})}}{\Delta\pt\times\Delta y\times L},
    \label{eq:xsection_def}
\end{equation}
where $L$ is the integrated luminosity,
$\Delta\pt$ and $\Delta y$ are the
interval sizes in terms of dimuon transverse momentum and centre-of-mass rapidity respectively,
and $N_{\mathcal{O}(n\text{S})}$ is the observed quarkonium yield in the kinematic interval under study, 
extracted from fits and corrected for acceptance, trigger and reconstruction efficiencies.
The total correction weight assigned to each selected dimuon candidate is given by:

\begin{equation}
        w_{\mathrm{total}}^{-1} = \mathcal{A}(\mathcal{O}(n\text{S})) \cdot \varepsilon_{\mathrm{reco}}\cdot\varepsilon_{\mathrm{trig}},
\label{eq:total_weight}
\end{equation}
where $\mathcal{A}(\mathcal{O}(n\text{S}))$ is the acceptance of the dimuon system for one of the five quarkonium states, 
$\varepsilon_{\mathrm{reco}}$ is the dimuon reconstruction efficiency
and $\varepsilon_{\mathrm{trig}}$ is the trigger efficiency.

\subsection{Acceptance}
\label{subsec:accep}

The acceptance of quarkonium decays into muon pairs is defined as
the probability of both muons from the decay falling in the fiducial region ($\pt(\mu^{\pm}) > 4~\GeV, |\eta(\mu^{\pm})|<2.4$).
The acceptance depends on transverse momentum, rapidity, invariant mass and the spin-alignment of the quarkonium state. 
The invariant mass of each state is taken to be the generator-level mass. 
Previous measurements in $pp$ collisions~\cite{CMS-BPH-11-023,CMS-BPH-13-003,LHCb_Jpsi_polarization} 
indicate that decays of quarkonia produced at LHC energies are consistent with the assumption that they are unpolarised.
Based on this assumption, and with a further assumption that the nuclear medium does not modify the average polarisation of produced quarkonia,
all quarkonium states in both the \pPb and $pp$ collisions are considered to be produced unpolarised in this paper.
The acceptance, $\mathcal{A}(\mathcal{O}(n\text{S}))$, for each of $\Jpsi$, $\psip$, and $\Upsins$ ($n = 1, 2, 3$) 
as a function of quarkonium transverse momentum and $|y|$ 
is calculated using generator-level MC, applying cuts on the $\pt$ and $\eta$ of the muons to emulate the fiducial volume as described in Refs.~\cite{BPHY-2010-01, BPHY-2011-06}.
The reconstructed dimuon transverse momentum, $\pt^{\mu\mu}$, is used for obtaining the acceptance correction for a given event.
However, the reconstructed dimuon transverse momentum and the quarkonium transverse momentum could be different due to final-state radiation from muons.
Corrections for final-state radiation are obtained by comparing acceptances calculated from 
generator-level muons with those after full detector simulation. 
The final-state radiation corrections as a function of $\pt^{\mu\mu}$ are applied to the acceptance corrections.
The correction factors are different for charmonium and bottomonium states but are the same for ground and excited states.
Finally, the same correction factors are used in \pp and \pPb data.

\subsection{Muon reconstruction and trigger efficiency}

The single muon reconstruction efficiency in \pPb data is determined directly from data 
using $\Jmm$ tag-and-probe method as used in Ref.~\cite{HION-2013-07,PERF-2014-05}, 
in which the tag muon is required to match with the trigger used to select the sample such that the probe muon is unbiased from the sample selection trigger, 
and the purity of the probe is guaranteed by background subtraction based on $\Jmm$ decay.
The dimuon trigger efficiency in \pPb data is factorised 
into single-muon trigger efficiencies for reconstructed muons at the L1 and HLT,  
with the correlation between the L1 and HLT trigger algorithms taken into account.
The single-muon trigger efficiencies are obtained from data, 
based on $\Jmm$ tag-and-probe method as described in Ref.~\cite{HION-2013-07},
in intervals of $\pt(\mu)$ and $q\times\eta(\mu)$, where $q$ is the charge of the muon.

For the $pp$ data, the same $\Jmm$ tag-and-probe technique as for \pPb data is used to determine single muon reconstruction and trigger efficiencies.
The dimuon trigger efficiency in the $pp$ data consists of two components.
The first part represents the trigger efficiency for a single
muon in intervals of $\pt(\mu)$ and $q\times\eta(\mu)$.
The second part is a dimuon correction term to account for reductions
in the trigger efficiency due to close-by muon pairs identified as single muon candidates at L1.
The dimuon correction term, which is determined separately for charmonium and bottomonium candidates, 
also accounts for inefficiency due to the vertex-quality requirement and opposite-sign charge requirement on the two online candidates. 
The efficiency and the dimuon correction term obtained from the MC simulation are used to correct data in order to suppress the statistical fluctuations of measured corrections. 
The measured average single-muon trigger efficiency is about $80\%$ ($95\%$) in the range $|\eta(\mu)| < 1.05$ ($1.05 < |\eta(\mu)| < 2.4$).
In addition to the main corrections derived from simulation, data-to-simulation scale factors, which are simple linear factors to account for the differences between data and MC simulation, are also applied.
The resulting scale factor is found to be about $92\%$ in the range $|\eta(\mu)| < 1.05$ 
and about $98\%$ in the range $1.05 < |\eta(\mu)| < 2.4$ without apparent $\pt$ dependence in both $\eta$ regions.

\subsection{Yield extraction}

\emph{Charmonium}

The charmonium yield determination decomposes the yields into two sources of muon pairs referred to as "prompt" and "non-prompt".
The prompt $\jpsi$ and $\psip$ signal originates from the strong production of short-lived particles,
including feed-down from other short-lived charmonium states,
while non-prompt refers to $\jpsi$ and $\psip$ mesons which are the decay products of $b$-hadrons.
To distinguish between these prompt and non-prompt processes, the
pseudo-proper lifetime, $\tau_{\mu\mu} = (L_{xy}m_{\mu\mu})/\pt^{\mu\mu}$, is used.
The transverse displacement, $L_{xy}$,
is the distance of the dimuon secondary vertex from the primary vertex along the dimuon momentum direction in the transverse plane.
Two-dimensional unbinned maximum-likelihood fits, as used in a previous ATLAS measurement~\cite{BPHY-2012-02}, 
are performed on weighted distributions of the dimuon invariant mass ($m_{\mu\mu}$)
and pseudo-proper lifetime ($\tau_{\mu\mu}$) to extract prompt and non-prompt signal yields, in intervals of $\pt^{\mu\mu}$, rapidity and centrality.
The event weight is given by Eq. (\ref{eq:total_weight}).
To obtain the acceptance corrections, $\Jpsi$ acceptance is applied to events with 
$m_{\mu\mu} < 3.2~\GeV$, $\psip$ acceptance is applied to events with $m_{\mu\mu} > 3.5~\GeV$ and 
a linear interpolation of the two acceptances is used for events with $3.2 < m_{\mu\mu} < 3.5~\GeV$.
Each interval of $\pt^{\mu\mu}$, rapidity and centrality is fitted independently in the RooFit framework~\cite{Roofit:2003}.
The two-dimensional probability density function (PDF) in $m_{\mu\mu}$ and $\tau_{\mu\mu}$ for the fit model is defined as:

\begin{equation*}
	\text{PDF}(m_{\mu\mu},\tau_{\mu\mu}) = \sum^7_{i=1}\kappa_i f_i(m_{\mu\mu}) \cdot h_i(\tau_{\mu\mu}) \otimes g(\tau_{\mu\mu}),
\label{eq:pdfdef}
\end{equation*}
where $\otimes$ implies a convolution, 
$\kappa_i$ is the normalisation factor of each component and
$g(\tau_{\mu\mu})$ is a double Gaussian $\tau_{\mu\mu}$ resolution function.
The two Gaussian components share a fixed mean at $\tau_{\mu\mu} = 0$.
One of the two widths in the resolution function is free,
while the other width is fixed at the first one multiplied by a constant factor,
determined from MC simulation. 
The relative fraction of the two Gaussian components is a free parameter.
The details of the seven components in the nominal fit model are summarised in Table~\ref{tab:pdf} and described below.

\begin{table}[t]
    \begin{center}
        \caption{Probability density functions for individual components in the central fit model used to extract the prompt and non-prompt contributions for charmonium signals and backgrounds.
        The composite PDF terms are defined as follows: $CB$ - Crystal Ball; $G$ - Gaussian; $E$ - Exponential; $F$ - constant distribution; $\delta$ - delta function.
        The parameter $\omega_i$ is the fraction of CB component in signal.}
        \begin{tabular}{ccccc}
            \hline
            $i$ & Type      	& Source    		& $f_i(m_{\mu\mu})$                                  						& $h_i(\tau_{\mu\mu})$ 	\\ \hline
            1 & $\jpsi$   		& Prompt         		& $\omega_1CB_1(m_{\mu\mu})+(1-\omega_1)G_1(m_{\mu\mu})$  	& $\delta(\tau_{\mu\mu})$ \\
            2 & $\jpsi$   		& Non-prompt        	& $\omega_1CB_1(m_{\mu\mu})+(1-\omega_1)G_1(m_{\mu\mu})$   & $E_1(\tau_{\mu\mu})$ 	\\
            3 & $\Ps$     		& Prompt         		& $\omega_2CB_2(m_{\mu\mu})+(1-\omega_2)G_2(m_{\mu\mu})$   & $\delta(\tau_{\mu\mu})$ \\
            4 & $\Ps$     		& Non-prompt        	& $\omega_2CB_2(m_{\mu\mu})+(1-\omega_2)G_2(m_{\mu\mu})$   & $E_2(\tau_{\mu\mu})$ 	\\ \hline
            5 & Background  	& Prompt         	   	& $F$                                    								& $\delta(\tau_{\mu\mu})$ \\
            6 & Background   	& Non-prompt        	& $E_3(m_{\mu\mu})$                                  					& $E_4(\tau_{\mu\mu})$ 	\\
            7 & Background    	& Non-prompt        	& $E_5(m_{\mu\mu})$                                  					& $E_6(|\tau_{\mu\mu}|)$ 	\\
            \hline
        \end{tabular}
        \label{tab:pdf}
    \end{center}
\end{table}

The signal charmonium line shape in $m_{\mu\mu}$ is described by the sum of a Crystal Ball
shape (CB)~\cite{paper:CB} and a single Gaussian function with a common mean.
The width parameter in the CB function is free,
while the Gaussian width is fixed with respect to the CB width 
by a constant factor motivated by the ratio of muon transverse momentum resolutions in different parts of the detector.
The rest of the parameters in the CB function are fixed to values obtained from MC simulation.
The mean and width of the $\psip$ are fixed to those of the $\Jpsi$
multiplied by a factor equal to the ratio of the measured masses of the $\psip$ and the $\Jpsi$~\cite{Agashe:2014kda}.
The relative fraction of the CB and Gaussian components is considered to be a free parameter, 
but one that is common to both the $\jpsi$ and $\psip$.
The prompt charmonium line shapes in $\tau_{\mu\mu}$ are described by a $\delta$ function
convolved with the resolution function $g(\tau_{\mu\mu})$,
whereas the non-prompt charmonium signals have pseudo-proper lifetime line shapes given by
an exponential function convolved with $g(\tau_{\mu\mu})$.

The background contribution contains one prompt component and two non-prompt components.
The prompt background is given by a $\delta$ function convolved with $g(\tau_{\mu\mu})$ in $\tau_{\mu\mu}$
and a constant distribution in $m_{\mu\mu}$.
One of the non-prompt background contributions is described by a single-sided exponential function convolved with $g(\tau_{\mu\mu})$ (for positive $\tau_{\mu\mu}$ only),
and the other non-prompt background contribution is described by a double-sided exponential function convolved with $g(\tau_{\mu\mu})$
accounting for misreconstructed or combinatoric dimuon pairs.
The two non-prompt backgrounds are parameterised as two independent exponential functions in $m_{\mu\mu}$.

\begin{figure}[!htb]
  \begin{center}
  
  \includegraphics[width=\fourfigsize]{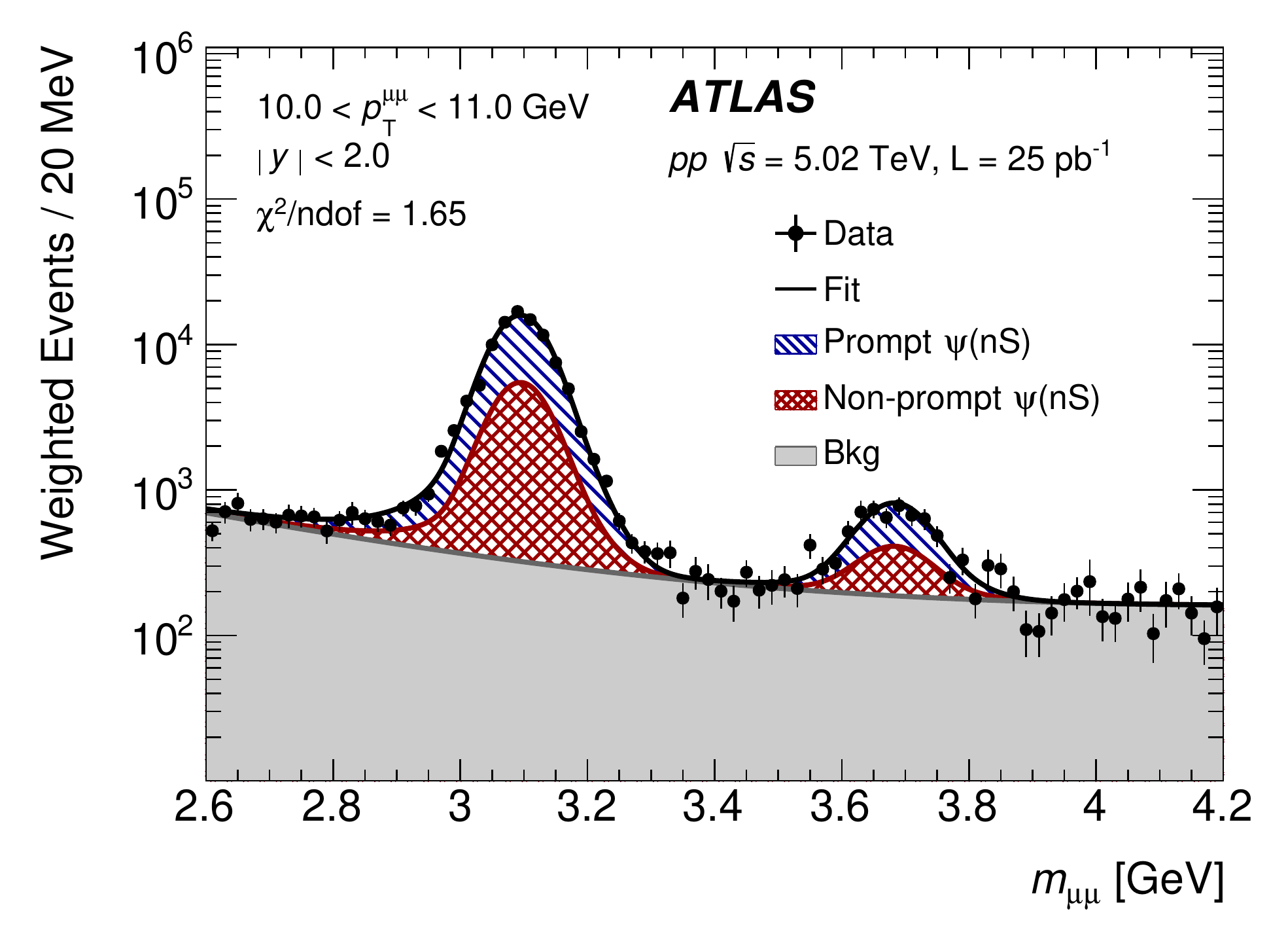}
  \includegraphics[width=\fourfigsize]{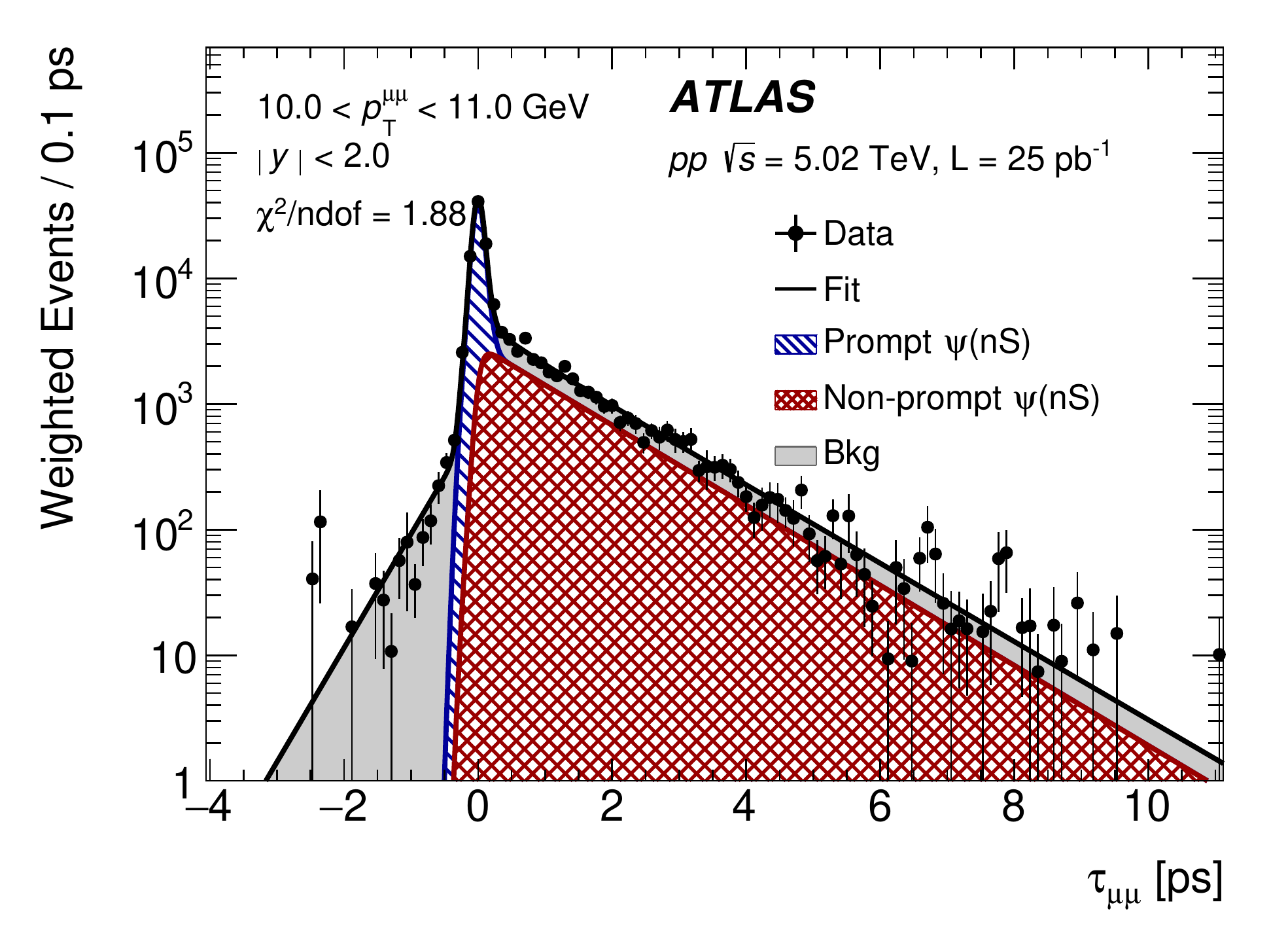}\\
  \includegraphics[width=\fourfigsize]{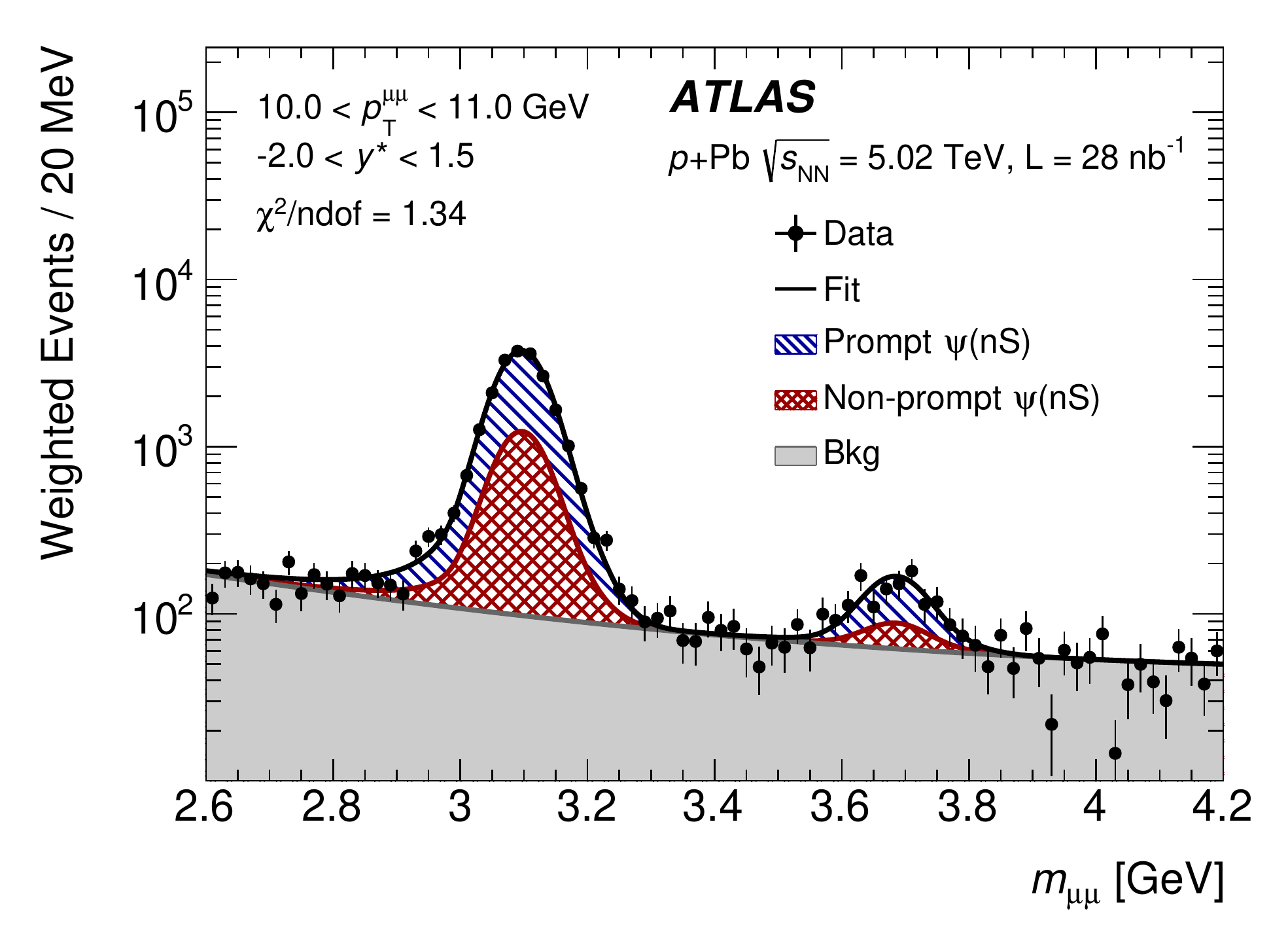}
  \includegraphics[width=\fourfigsize]{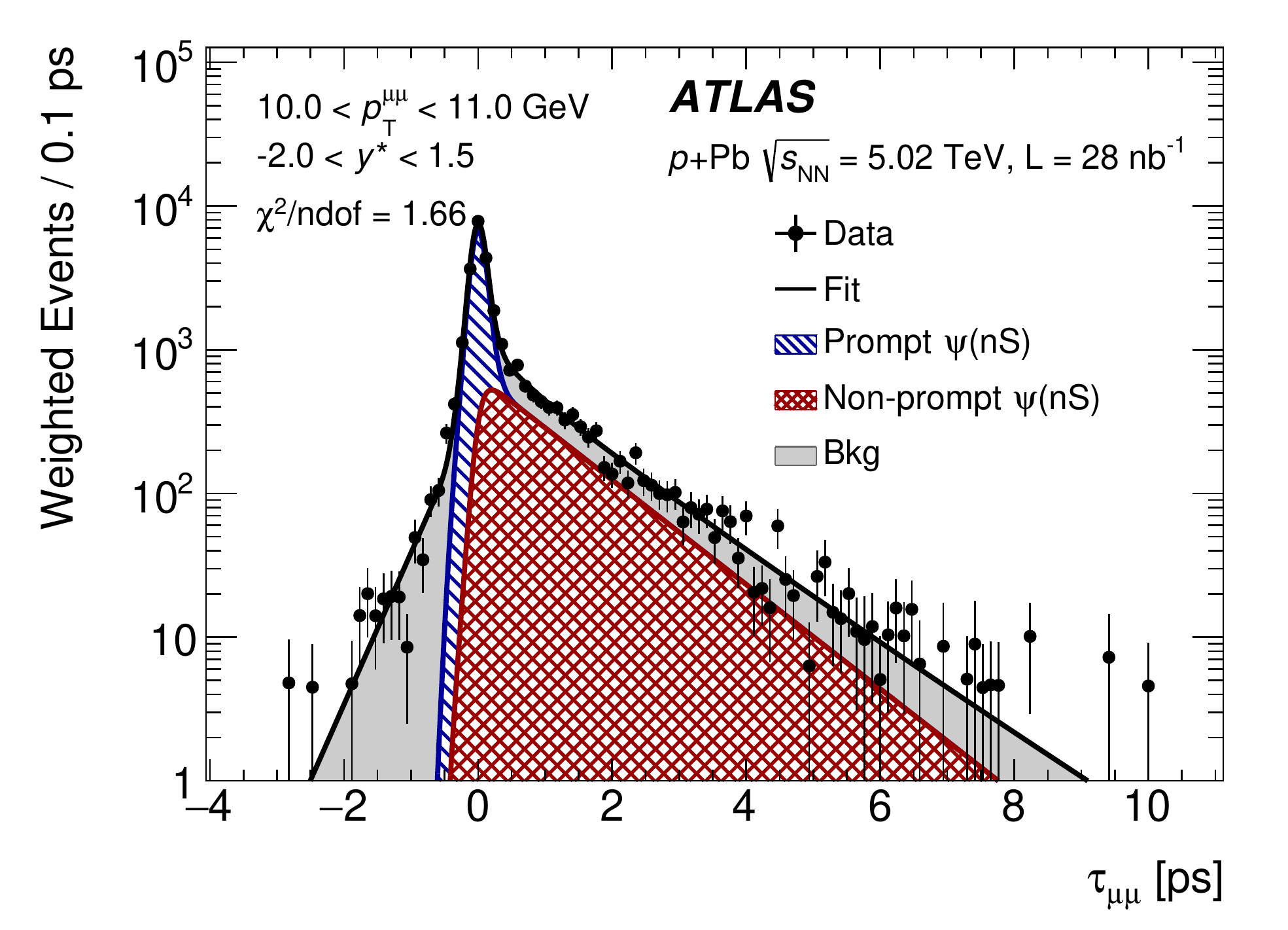}\\
  \caption{Projections of the charmonium fit results onto dimuon invariant mass $m_{\mu\mu}$ (left) 
  and pseudo-proper lifetime $\tau_{\mu\mu}$ (right)
  for $pp$ collisions at $\ppenergy$ (top) for the kinematic ranges $10 < \pt^{\mu\mu} < 11~\GeV$ and $|y| < 2.0$,
  and \pPb collisions at $\energy$ (bottom)  for the kinematic ranges $10 < \pt^{\mu\mu} < 11~\GeV$ and $-2.0 < y^* < 1.5$.
  The goodnesses of the invariant mass fits with $\text{ndof} = 63$
  and the pseudo-proper lifetime fits with $\text{ndof} = 153$
  are also presented.
  }
  \label{fig:2D_fit}
  \end{center}
\end{figure}

There are in total seventeen free parameters in the charmonium fit model.
The normalisation factor $\kappa_i $ of each component is extracted from each fit.
From these parameters, and the weighted sum of events, all measured values are calculated.
Figure~\ref{fig:2D_fit} shows examples of charmonium fit projections onto invariant mass and pseudo-proper lifetime axes.
The fit projections are shown for the total prompt signal, total non-prompt signal and total background contributions.

\emph{Bottomonium}

The yields of bottomonium states are obtained by performing unbinned maximum likelihood fits of the weighted invariant mass distribution,
in intervals of $\pt^{\mu\mu}$, rapidity and centrality.
Due to overlaps between the invariant mass peaks of different bottomonium states,
the linear acceptance interpolation used for the charmonium states is not appropriate to the bottomonium states.
Instead, each interval is fitted three times to extract the corrected yields of the three different bottomonium states, 
each time with the acceptance weight of one of the three states assigned to all candidates. 
Each of the fits in each interval of $\pt^{\mu\mu}$, rapidity and centrality is independent of all the others. 

\begin{figure}[!htb]
  \begin{center}
  \includegraphics[width=\fourfigsize]{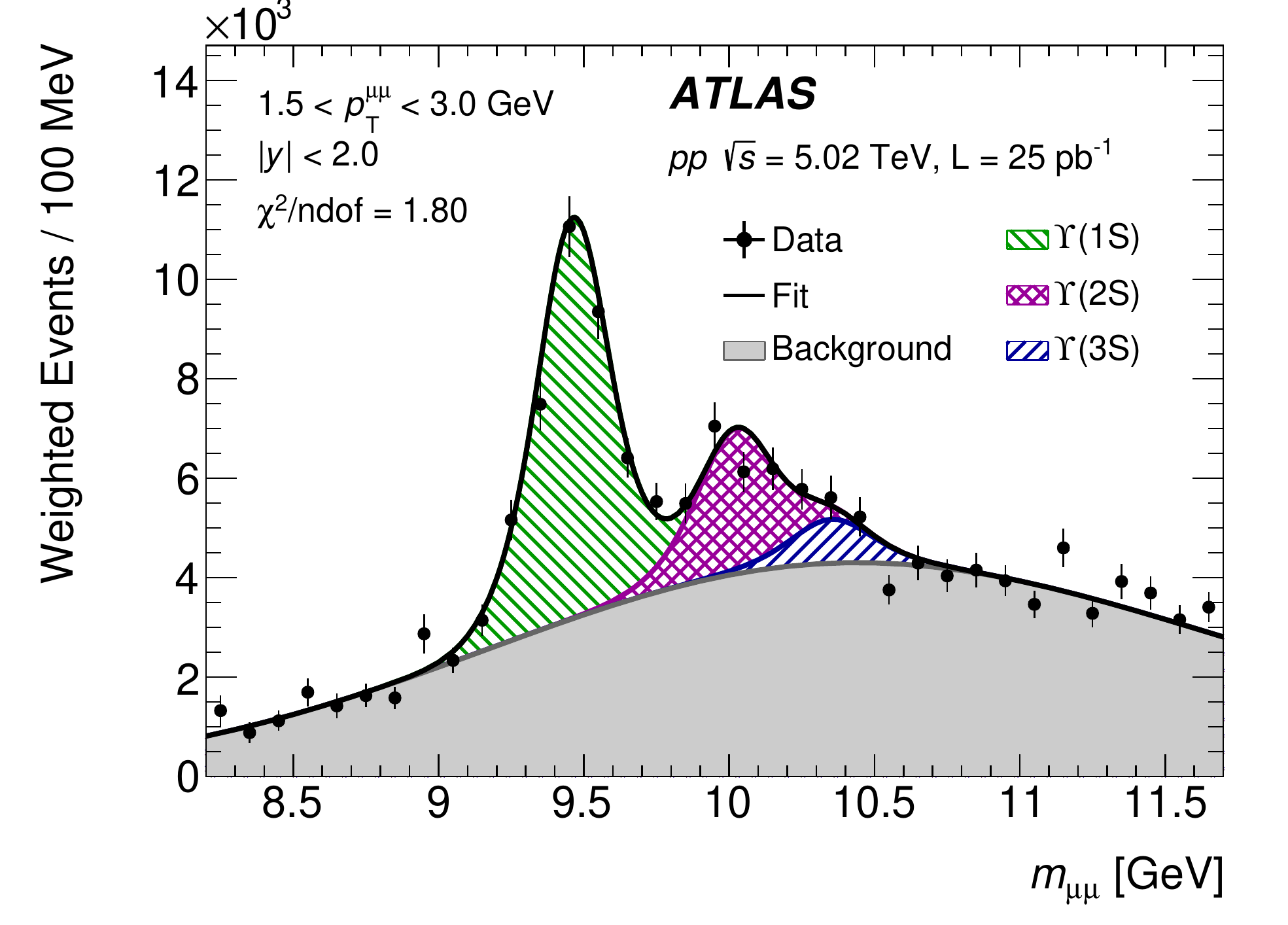}
  \includegraphics[width=\fourfigsize]{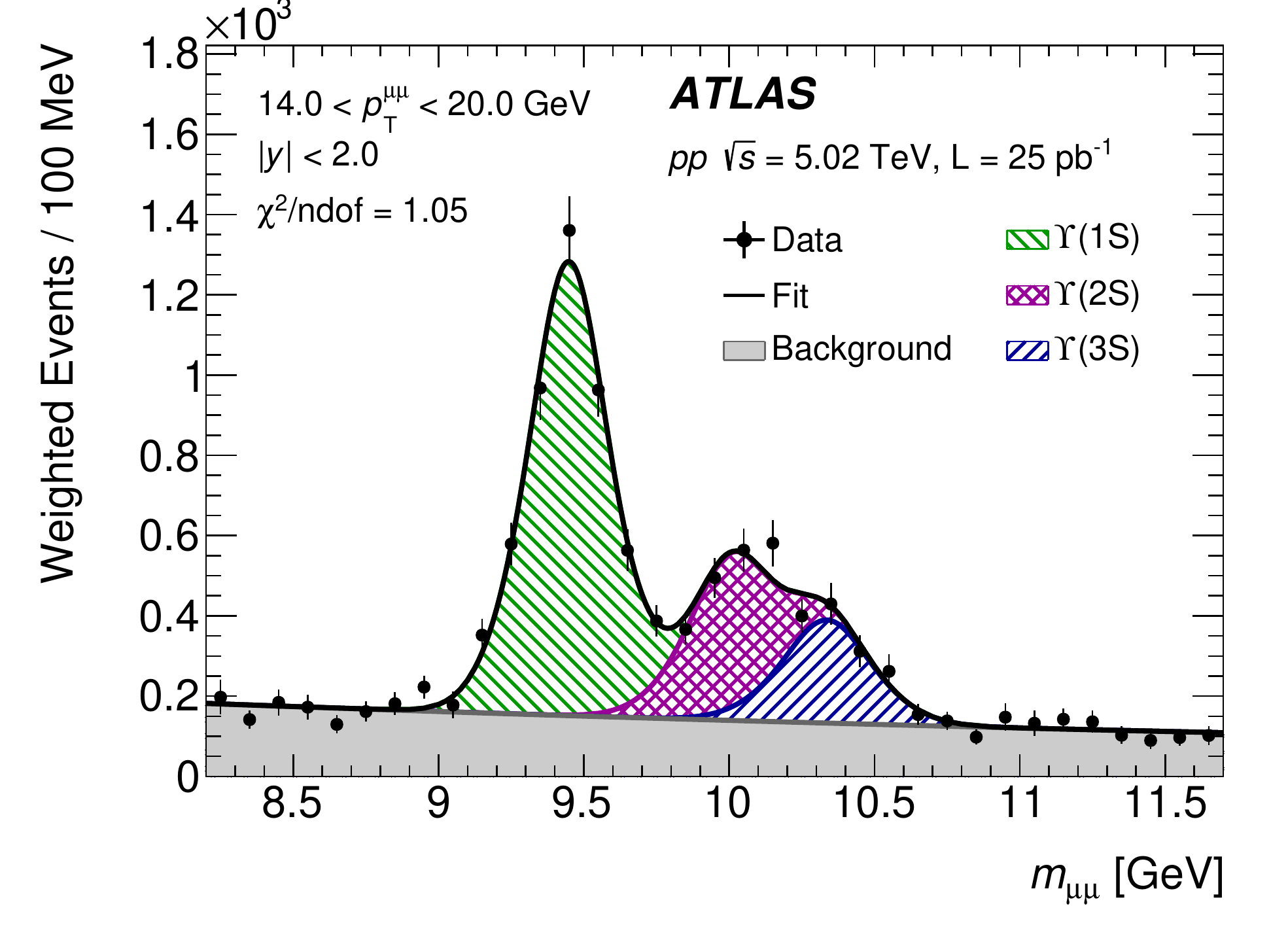}\\
  \includegraphics[width=\fourfigsize]{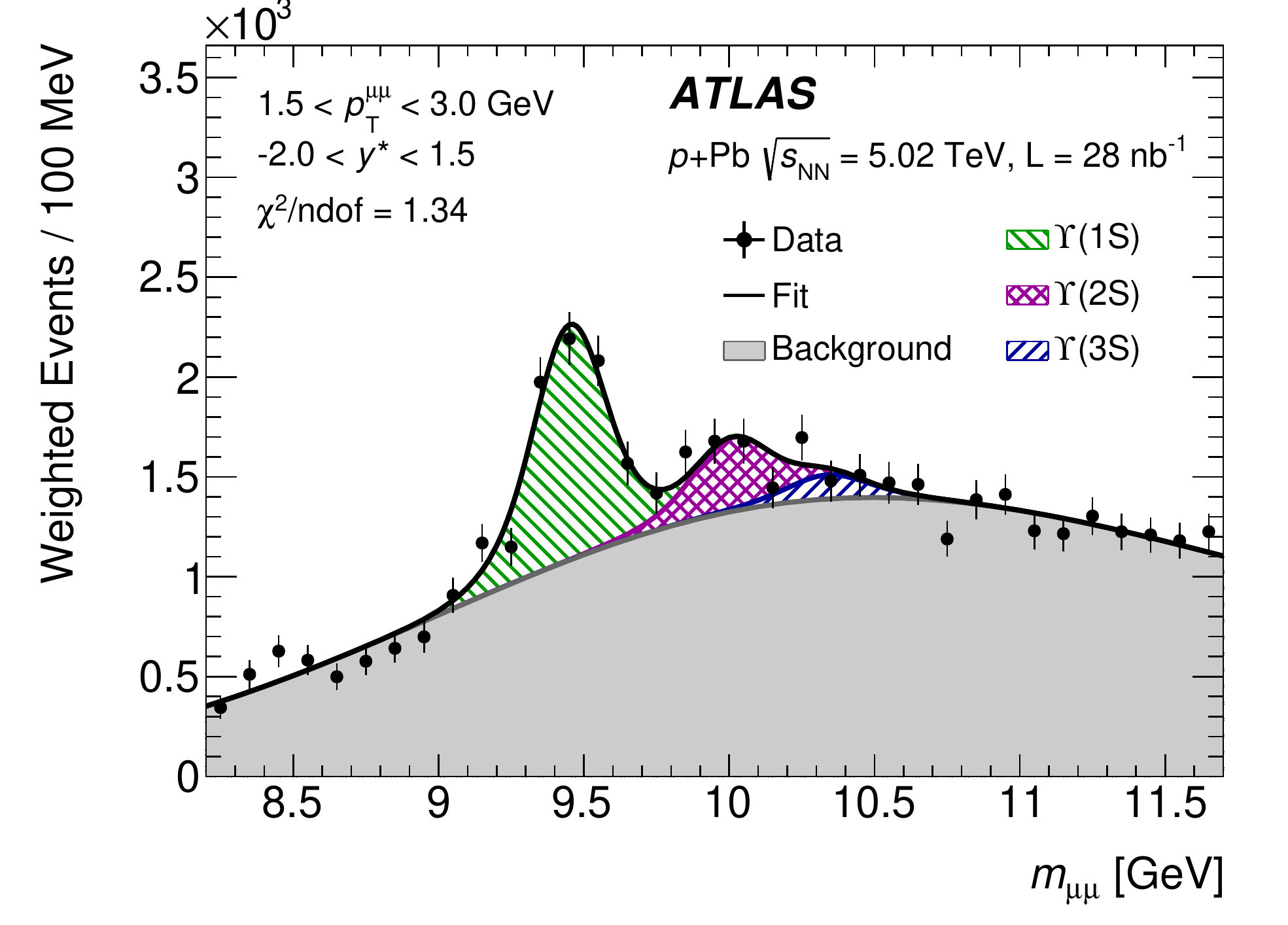}
  \includegraphics[width=\fourfigsize]{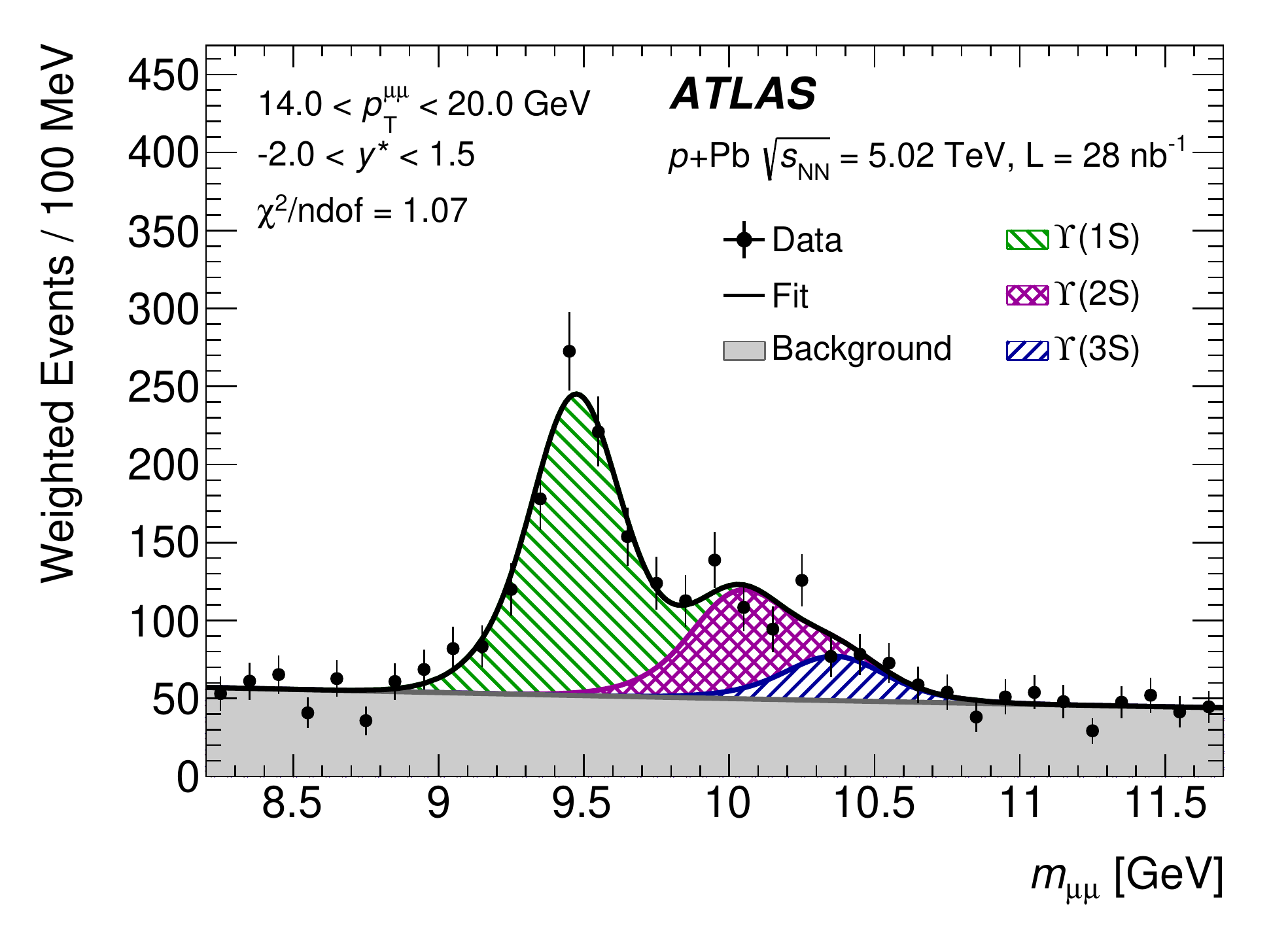}\\
  \caption{Bottomonium fit results in dimuon invariant mass $m_{\mu\mu}$
  for $pp$ collisions at $\ppenergy$ (top) and \pPb\ collisions at $\energy$ (bottom) for one typical low $\pt^{\mu\mu}$ interval of $1.5<\pt^{\mu\mu}<3~\GeV$ (left) and one high $\pt^{\mu\mu}$ interval of $14<\pt^{\mu\mu}<20~\GeV$ (right).
  For all the shown fits, $\Upsi$ acceptance weights are assigned.
  The goodness of the bottomonium fit with $\text{ndof} = 24$ is also presented.}
  \label{fig:fit_template}
  \end{center}
\end{figure}

The bottomonium signal invariant mass model is essentially the same as the charmonium model.
The mean and width of the $\Upsi$ is free, while the  means 
and widths of $\Upsii$ and $\Upsiii$ are fixed with respect to parameters of $\Upsi$
with a constant scaling factor equal to the $\Upsins$ to $\Upsi$ mass ratio taken from Ref.~\cite{Agashe:2014kda}.
The bottomonium background parameterisation varies with $\pt^{\mu\mu}$.
At low $\pt^{\mu\mu}$ ($\pt^{\mu\mu} < 6~\GeV$), and for all rapidity intervals, an error function multiplied by an exponential function is used to model 
the $m_{\mu\mu}$ turn-on effects
due to decreasing acceptance with decreasing invariant mass, which originates from the \pt selection
applied to each muon.
At low $\pt^{\mu\mu}$, the background model's parameters are constrained by using a background control sample.
The control sample is selected from dimuon events in which at least one of the muons has
a transverse impact parameter with respect to the primary vertex larger than $0.2~\mathrm{mm}$.
This criterion causes the control sample to be dominated by muon pairs from the decay of $b$-hadrons.
For candidates with higher $\pt^{\mu\mu}$, a second-order polynomial is used to describe the background contribution.
At low $\pt^{\mu\mu}$, the background model is first fitted to the control sample,
then the parameters of the error function are fixed at their fitted values,
and finally the full fit model with the constrained background is applied to the data sample.
Some selected bottomonium fits are shown in Figure~\ref{fig:fit_template}.

\section{Systematic uncertainties}
\label{sec:systematic}

The sources of systematic uncertainty in the quarkonium yields include 
acceptance, muon reconstruction and trigger efficiency corrections, the fit model parameterisation and bin migration corrections and the luminosity. 
For the ratio measurements the systematic uncertainties are assessed in the same manner as for the yields, 
except that in the ratios the correlated systematic uncertainties, such as the luminosity uncertainty, cancel out. 

\emph{Luminosity}

The uncertainty in integrated luminosity is $2.7\%$ ($5.4\%$) for 2013 \pPb\ (2015 $pp$) data-taking.
The luminosity calibration is 
based on data from dedicated beam-separation scans, also known as van der Meer scans, as described in Ref.~\cite{DAPR-2013-01}.

\emph{Acceptance}

A systematic uncertainty for the final-state radiation corrections is assigned
to cover the differences between correction factors obtained for ground and excited states of quarkonium 
and for different rapidity slices.
The systematic uncertainties fully cancel out in ratio measurements in the same datasets and between different datasets.

\emph{Muon reconstruction and trigger efficiencies in \pPb\ collisions}

The dominant source of uncertainty in the muon reconstruction and trigger efficiency in \pPb collisions is statistical.
Therefore, the uncertainty in each bin is treated as uncorrelated
and the corresponding uncertainties are propagated to the measured observables by using pseudo-experiments as in previous ATLAS measurements~\cite{BPHY-2012-02}.
For each pseudo-experiment a new efficiency map is created by varying independently the content of each bin according to a Gaussian distribution.
The mean and width parameters of the Gaussian distribution are respectively the value and uncertainty of the bin in the original map.
In each pseudo-experiment, the total weight is recalculated for each dimuon kinematic interval of the analysis.
A distribution of total weight is obtained from repeating pseudo-experiments for 200 times, which is sufficient to suppress the statistical fluctuation of the sample used in each experiment.
For each efficiency type, the RMS of the total weight distributions is assigned as the systematic uncertainty. 

An additional uncertainty of $1\%$ is applied to cover the small muon reconstruction inefficiency in the inner detector in \pPb\ collisions.
The dimuon trigger efficiency factorisation is tested in simulation,
and a bias of at most $4\%$ is found in yield observables.
The bias stems from the imperfect approximation of the correlation between trigger algorithms at different levels in the dimuon trigger factorisation.
An additional correlated uncertainty of $4\%$ is added to cover this bias.
This uncertainty is applied to quarkonium yields in \pPb\ collisions,
but is assumed to cancel in ratios measured in the same datasets.

\emph{Muon reconstruction and trigger efficiencies in $\pp$ collisions}

For the $pp$ measurements, the efficiency maps are determined from MC simulation and corrected with measured data-to-simulation scale factors as detailed in Ref.~\cite{PERF-2015-10}.
The statistical uncertainty associated with efficiency scale factor is evaluated using random replicas of the efficiency maps as for \pPb\
and the different sources of uncertainty described below are treated as correlated.
The systematic uncertainty in reconstruction efficiency is obtained by varying the signal and background models in the fits used to extract the efficiency in data,
and taking the difference between the reconstruction efficiency calculated using generator-level information and the value obtained with the tag-and-probe method in MC simulation.
An additional $1\%$ correlated uncertainty is added to cover a systematic variation due to a small misalignment in the ID.
For the trigger efficiency, the following variations of the analysis are studied and the effects are combined in quadrature:
\begin{itemize}
    \item variations of signal and background fit model used to extract the data efficiency;
    \item variations of the matching criteria between a muon and a trigger element;
    \item using dimuon correction terms determined at positive (or negative) rapidity for whole rapidity range.
\end{itemize}
A test of the approximation of muon--muon correlation at L1 in the \pp dimuon trigger factorisation in MC simulation
results in a bias of at most $4\%$, which is the same size as the factorisation bias of the \pPb trigger but with totally different origins.
An additional $4\%$ correlated uncertainty is added to quarkonium yields to cover 
the bias.
This uncertainty cancels out in ratio observables that are measured in the same datasets.

\emph{Bin migrations}

Corrections due to bin migration factors were evaluated in Refs.~\cite{BPHY-2012-02, BPHY-2011-06} 
and are determined to be less than $0.5\%$ of the measured values.  
For this reason, bin migration correction factors and their uncertainties are neglected in this analysis.

\emph{Charmonium fit}

The uncertainty from the signal and background line shapes is estimated from variations of the fit model.
To remove the statistical component, each variation is repeated with pseudo-experiments,
generated using the bootstrap method~\cite{bootstrap}. 
First, for each toy sample, every event from the original data is filled into the toy sample $n$ times,
where $n$ is a random integer obtained from a Poisson distribution with a mean of one.
Then the central model and a set of `variation' models are fitted to the toy sample,
and all measured quantities are recalculated.
The difference between the central model and a given variation model is extracted and recorded.
After repeating the pseudo-experiment 100 times,
the systematic uncertainty due to the line shape is defined as the mean difference of a given variation model from the nominal model.
Up to ten variation models are considered for the charmonium fit model, categorised into four groups:

\begin{itemize}
    \item \emph{Signal tail due to final-state radiation}.
    Evaluated by replacing the CB plus Gaussian model with a double Gaussian function, and varying the tail parameters of the CB model, which are originally fixed.
    \item \emph{Pseudo-proper lifetime resolution}.
    Evaluated by replacing the double Gaussian function with a single Gaussian function to model pseudo-proper lifetime resolution.
    \item \emph{Signal pseudo-proper lifetime shape}.
    Evaluated by using a double exponential function to describe the pseudo-proper lifetime distribution of the signal.
    \item \emph{Background mass shapes}.
    Evaluated by using a second-order Chebyshev polynomial to describe the prompt, non-prompt and double-sided background terms.
\end{itemize}

The total systematic uncertainty from the line shape fit is determined by combining the maximum variation found in each of the four groups in quadrature.
In order to estimate the possible bias introduced by the line shape assumptions in the nominal fit model parameterisation,
the nominal model is tested using the $\Jmm$ MC sample in which random numbers of prompt and non-prompt $\Jpsi$ are mixed.
About $1\%$ difference between the random input and fit model output is found for the yield and non-prompt fraction in the MC test.
An additional systematic uncertainty of $1\%$ is assigned to charmonium yields and non-prompt fractions to cover the nominal fit model parameterisation bias.
This uncertainty cancels out in the $\psip$ to $\jpsi$ yield ratio.

\emph{Bottomonium fit}

The systematic uncertainty from varying the fit model is estimated based on the same method
as used for the charmonium fit uncertainty,
and there are six variation models for bottomonium categorised into three groups:

\begin{itemize}
    \item \emph{Signal resolution}.
    Evaluated by replacing the CB plus Gaussian model with a single CB function and a triple Gaussian function,
    and varying the constant width scaling term between the CB function and the Gaussian function.
    \item \emph{Signal tail due to final-state radiation}.
    Evaluated by replacing the CB plus Gaussian model with a double Gaussian function,
    and treating the tail parameters in the CB function as free parameters.
    \item \emph{Background shapes}.
    Evaluated by replacing the low $\pt$ background distribution with a fourth-order Chebyshev polynomial,
    and replacing the high $\pt$ distribution by an exponential function or a second-order Chebyshev polynomial.
\end{itemize}

The total systematic uncertainty from the line shape fit is given by combining the maximum variation found in each of the three groups in quadrature.
An additional systematic uncertainty of $1.5\%$ is assigned to $\Upsi$ yields and $2\%$ for $\Upsins$ yields and $\Upsins$ to $\Upsi$ ratios ($n=2,3$),
in order to cover the bias of the nominal model found in MC tests similar to those for the charmonium fit model.

Table~\ref{Tab:syst} summaries the systematic uncertainties in the ground-state and excited-state yields and their ratio.
The dominant sources of systematic uncertainty for the yields are the fit model and muon trigger efficiency.
The ranges of uncertainties shown in the table indicate the minimum and maximum values found in all $\pt^{\mu\mu}$, rapidity and centrality intervals.  
The large range of bottomonium fit systematic uncertainty is due to the different modelling of the background at low $\pt^{\mu\mu}$ ($\pt^{\mu\mu} < 6~\GeV$) and high $\pt^{\mu\mu}$.
The systematic uncertainty from fit model variations is much larger at low $\pt^{\mu\mu}$ than at high $\pt^{\mu\mu}$.
For the ratios measured in the same datasets, most sources of systematic uncertainty including the trigger efficiency largely cancel out.

\begin{table}[t]
    \centering
     \caption{
    Summary of systematic uncertainties in the charmonium and bottomonium ground-state and excited-state yields and their ratio.
    The ranges of uncertainties indicate the minimum and maximum values found in all kinematic slices. Symbol "$-$" in the ratio observable column indicates the uncertainty fully cancels out.}
    \begin{tabular}{l|l|r|r|r}
        \hline
        Collision type	&Sources                	& Ground-state         & Excited-state    &Ratio           \\
                                	&                  		&yield $[\%]$            &yield $[\%]$       &$[\%]$          	\\ \hline
        \multirow{4}{*}{\pPb\ collisions}
                                &Luminosity             	& $2.7$                     & $2.7$           	& $-$              	\\
                                &Acceptance		& $1$--$4$		& $1$--$4$	& $-$		\\		
                                &Muon reco.             	& $1$--$2$                & $1$--$2$       & $< 1$          	\\
                                &Muon trigger           	& $4$--$5$                & $4$--$5$       & $< 1$          	\\
                                &Charmonium fit      	& $2$--$5$                & $4$--$10$     & $7$--$15$     \\
                                &Bottomonium fit     	& $2$--$15$              & $2$--$15$     & $5$--$12$     \\ \hline
        \multirow{4}{*}{$pp$ collisions}
                                &Luminosity             	& $5.4$                     & $5.4$           	& $-$              	\\
                                &Acceptance		& $1$--$4$		& $1$--$4$	& $-$		\\                           
                                &Muon reco.             	& $1$--$5$                & $1$--$5$       & $< 1$          	\\
                                &Muon trigger           	& $5$--$7$                & $5$--$7$       & $< 1$          	\\
                                &Charmonium fit       	& $2$--$7$                & $4$--$10$     & $7$--$11$        \\
                                &Bottomonium fit      	& $1$--$15$              & $2$--$15$  	& $5$--$12$      \\ \hline
    \end{tabular}
    \label{Tab:syst}
\end{table}

The luminosity systematic uncertainties in \pp and \pPb collisions are considered to be totally uncorrelated.
The acceptance systematic uncertainties in \pp and \pPb collisions are fully correlated. 
The reconstruction efficiency systematic uncertainties are treated as uncorrelated due to different muon selection criteria in \pp and \pPb collisions.
The uncertainties in the trigger efficiencies are also treated as uncorrelated 
since different efficiency determination strategies are used in \pp and \pPb collisions 
and the factorisation biases originate from different types of trigger correlations.
The fit model variation systematic uncertainties are found to be partially correlated 
and their effects on $\rpb$ and $\rhopb$
are determined by studying these ratios obtained from simultaneous fits to \pp and \pPb collision data for each variation.

\section{Results}
\label{sec:result}

\subsection{Production cross sections}

Following the yield correction and signal extraction,
the cross sections of five quarkonium states are measured differentially in transverse momentum\footnote{The transverse momentum of the quarkonium state is denoted as \pt 
in the rest of the paper.} and rapidity,
as described in Eq. (\ref{eq:xsection_def}).

\begin{figure}[!htb]
  \begin{center} 
  \includegraphics[width=\fourfigsize]{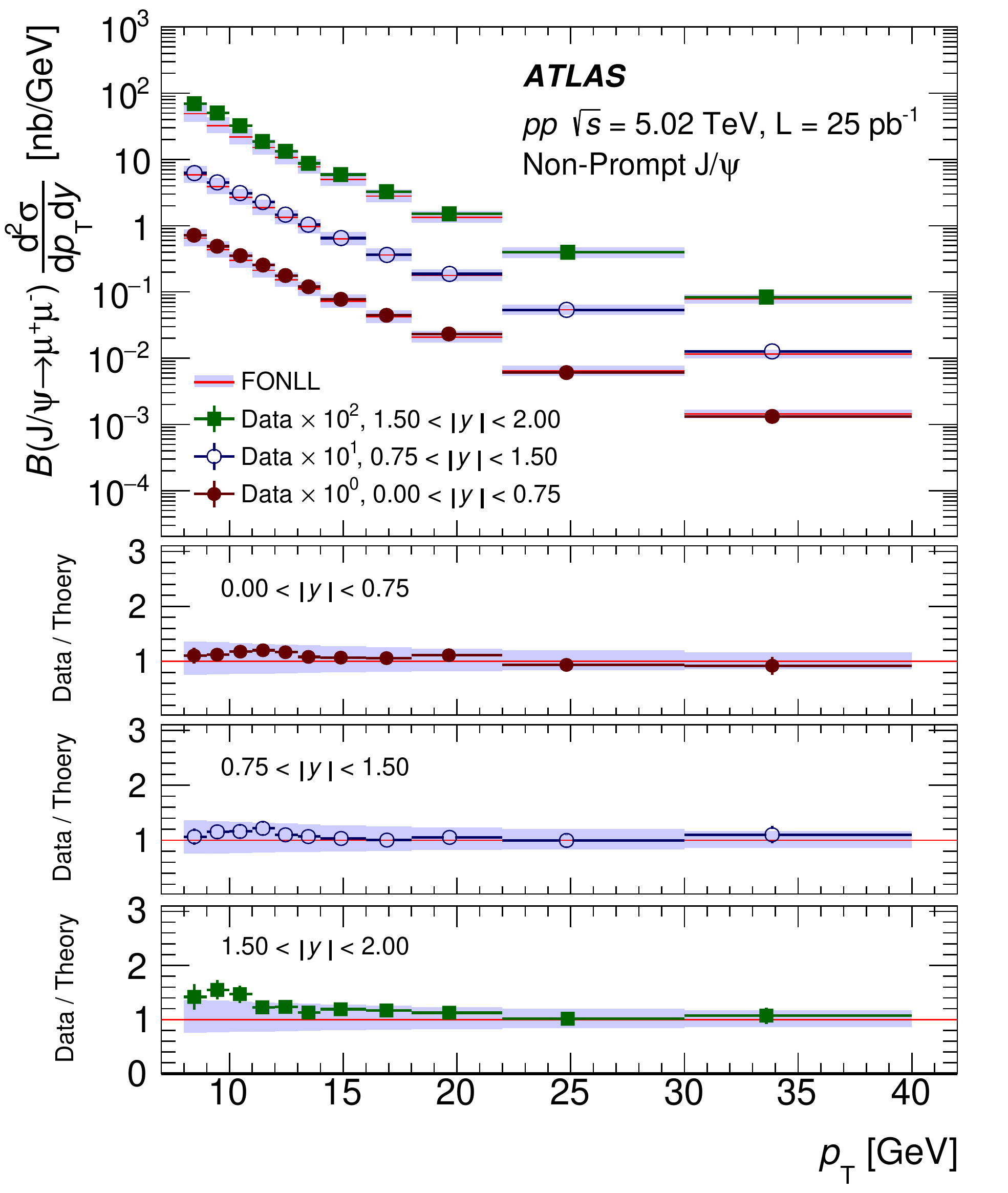}
  \includegraphics[width=\fourfigsize]{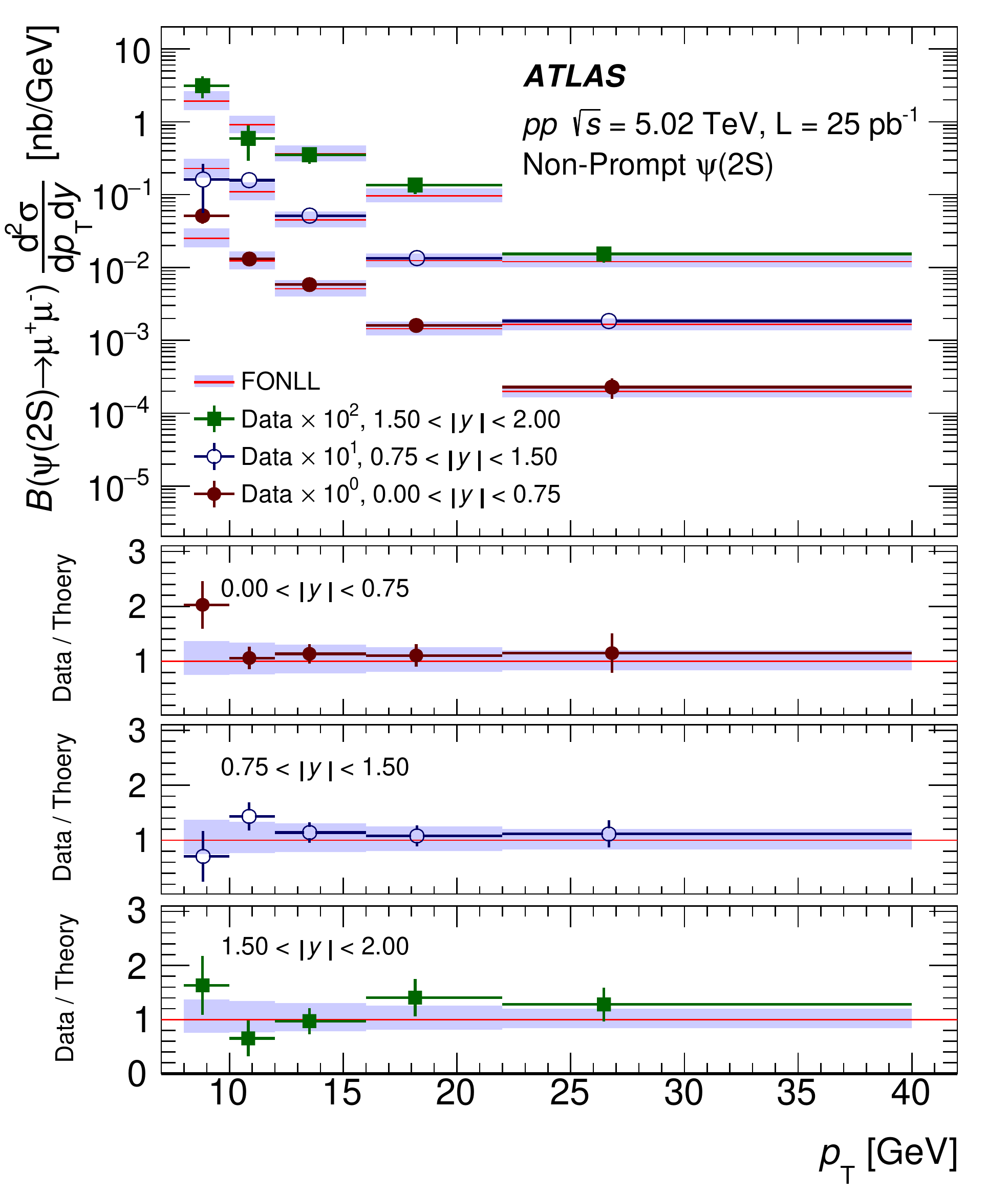}\\
  \caption{
  The differential non-prompt production cross section times dimuon branching fraction of $\Jpsi$ (left) and $\psip$ (right)
  as a function of transverse momentum $\pt$ for three intervals of rapidity $y$ in $pp$ collisions at $5.02~\TeV$.
  For each increasing rapidity slice, an additional scaling factor of 10 is applied to the plotted points for visual clarity.
  The horizontal position of each data point indicates the mean of the weighted $\pt$ distribution.
  The horizontal bars represent the range of $\pt$ for the bin,
  and the vertical error bars correspond to the combined statistical and systematic uncertainties.
  The FONLL theory predictions (see text) are also shown,
  and the error bands in the prediction correspond to the combined factorisation scale, quark mass and parton distribution functions uncertainties.
  }
  \label{fig:FONLL_pp}
  \end{center}
\end{figure}

The results for non-prompt $\Jpsi$ and $\psip$ cross sections in $pp$ collisions at $\ppenergy$ compared to fixed-order next-to-leading-logarithm (FONLL) predictions~\cite{Cacciari:2012ny}
are shown in intervals of $\pt$ for different rapidity slices in Figure~\ref{fig:FONLL_pp}.
The FONLL uncertainties include renormalisation and factorisation scale variations, charm quark mass and parton distribution functions uncertainties as detailed in Ref.~\cite{Cacciari:2012ny}.
The measured non-prompt charmonium production cross sections agree with the FONLL predictions within uncertainties over the measured $\pt$ range.

\begin{figure}[!htb]
  \begin{center}
  \includegraphics[width=\fourfigsize]{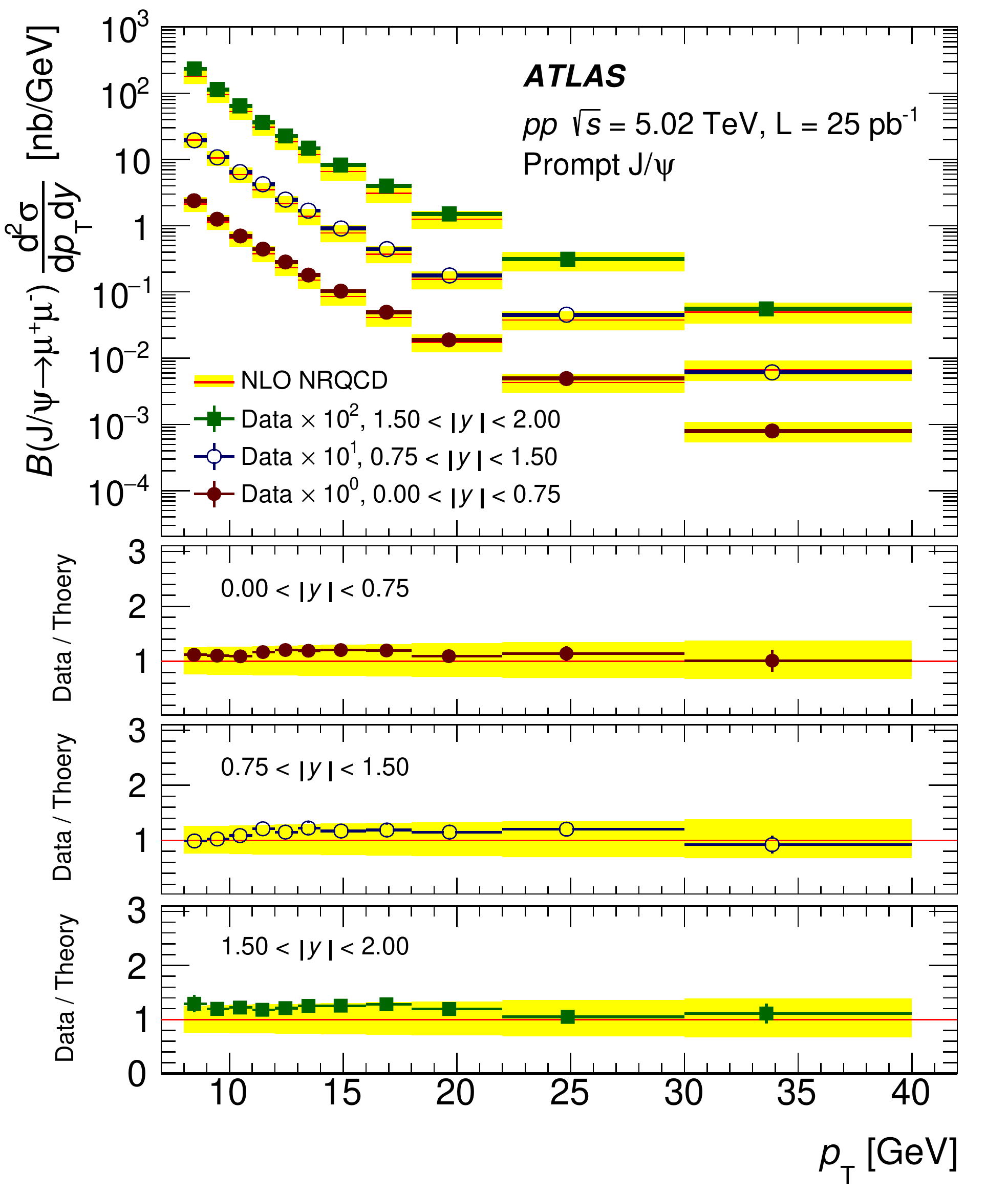}
  \includegraphics[width=\fourfigsize]{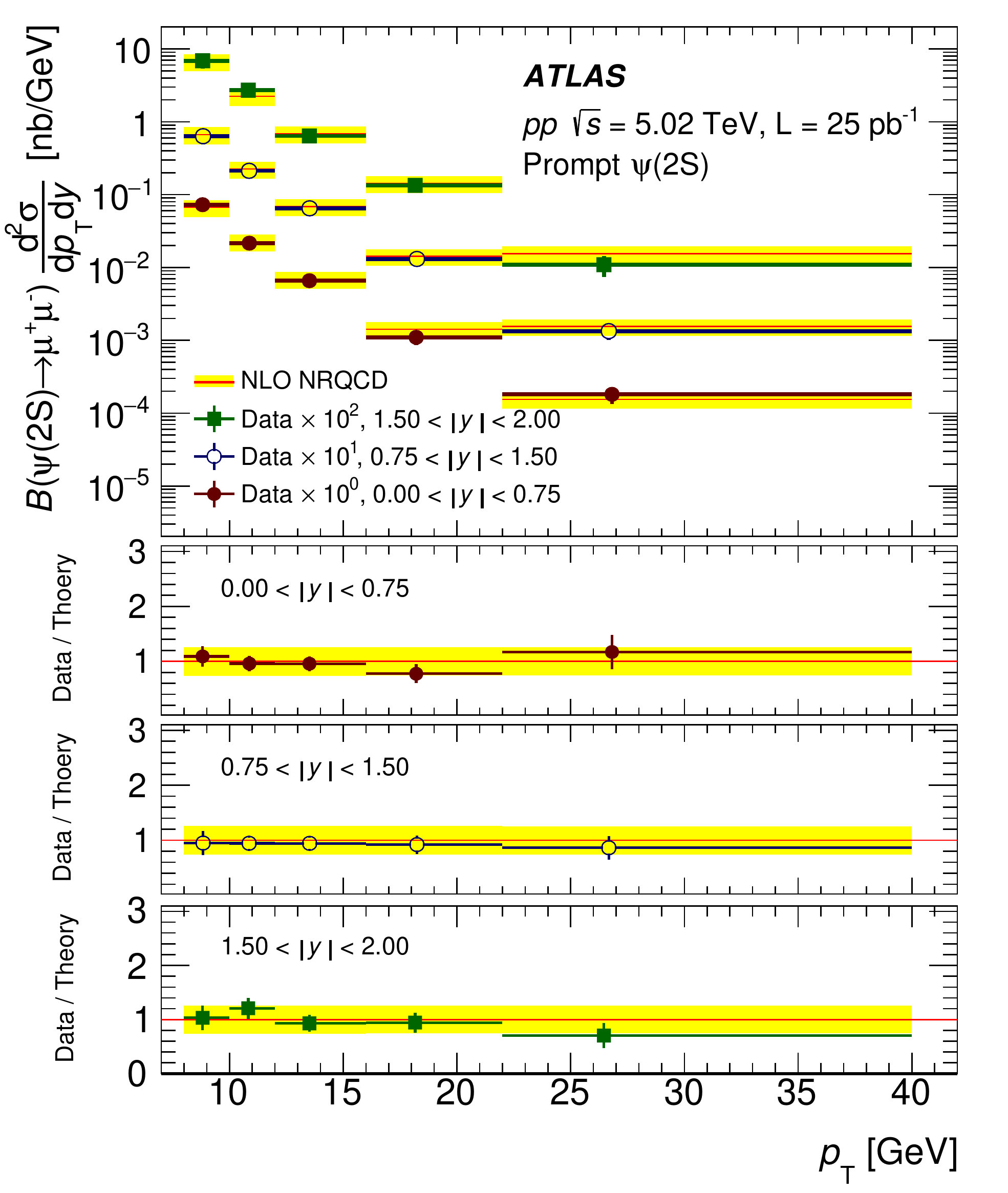}\\
  \caption{
  The differential prompt production cross section times dimuon branching fraction of $\Jpsi$ (left) and $\psip$ (right)
  as a function of transverse momentum $\pt$ for three intervals of rapidity $y$ in $pp$ collisions at $5.02~\TeV$.
  For each increasing rapidity slice, an additional scaling factor of 10 is applied to the plotted points for visual clarity.
  The horizontal position of each data point indicates the mean of the weighted $\pt$ distribution.
  The horizontal bars represent the range of $\pt$ for the bin,
  and the vertical error bars correspond to the combined statistical and systematic uncertainties.
  The NRQCD theory predictions (see text) are also shown,
  and the error bands in the prediction correspond to the combined scale, quark mass and LDMEs uncertainties.
  }
  \label{fig:NLO_pp}
  \end{center}
\end{figure}

\begin{figure}[!htb]
  \begin{center}  
  \includegraphics[width=\fourfigsize]{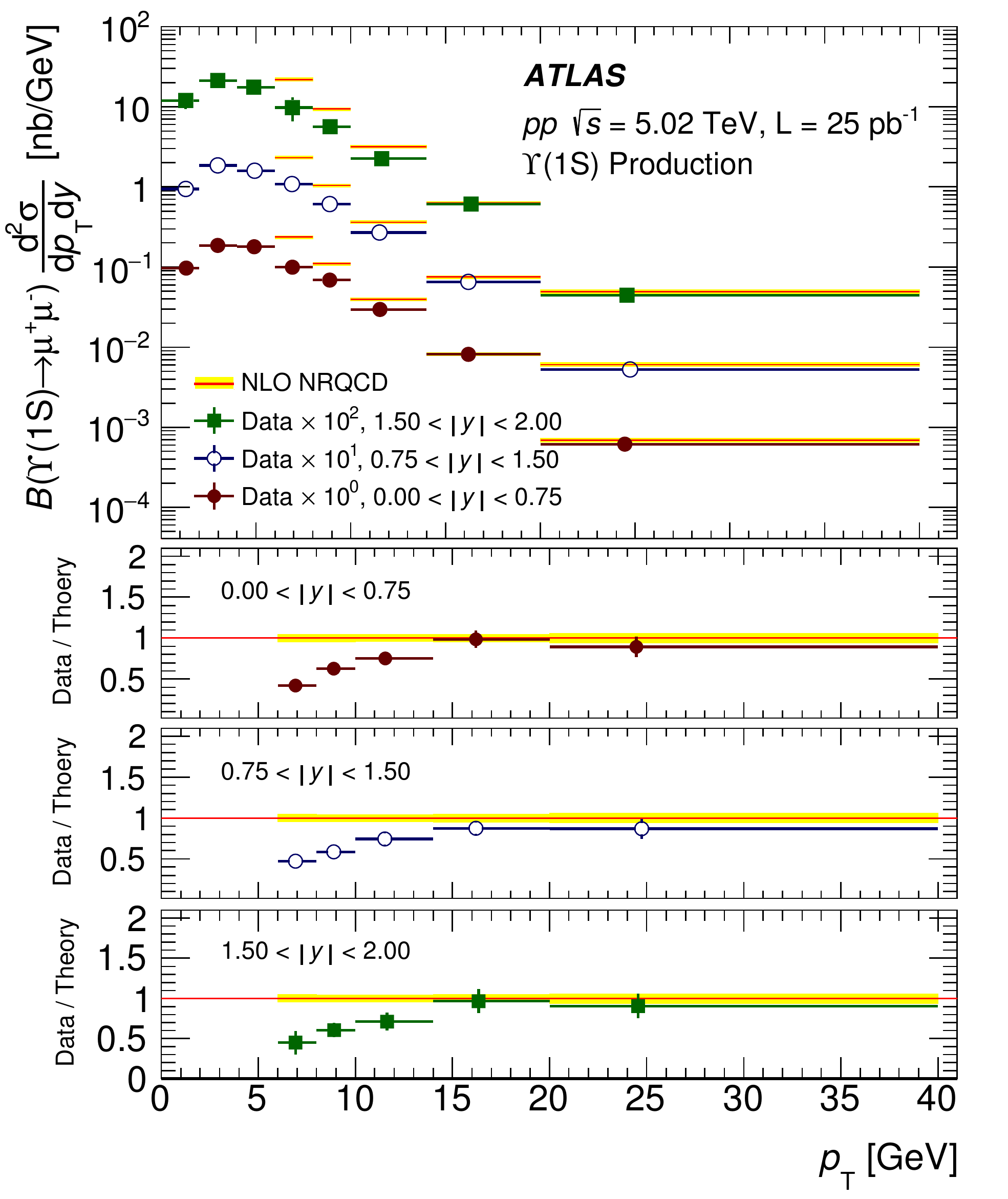}\\
  \caption{
  The differential production cross section times dimuon branching fraction of $\Upsi$
  as a function of transverse momentum $\pt$ for three intervals of rapidity $y$ in $pp$ collisions at $5.02~\TeV$.
  For each increasing rapidity slice, an additional scaling factor of 10 is applied to the plotted points for visual clarity.
  The horizontal position of each data point indicates the mean of the weighted $\pt$ distribution.
  The horizontal bars represent the range of $\pt$ for the bin,
  and the vertical error bars correspond to the combined statistical and systematic uncertainties.
  The NRQCD theory predictions (see text) are also shown,
  and the error bands in the prediction correspond to the combined scale, quark mass and LDMEs uncertainties.
  }
  \label{fig:Upsi1S_Xsec_pp}
  \end{center}
\end{figure}

\begin{figure}[!htb]
  \begin{center}
  \includegraphics[width=\fourfigsize]{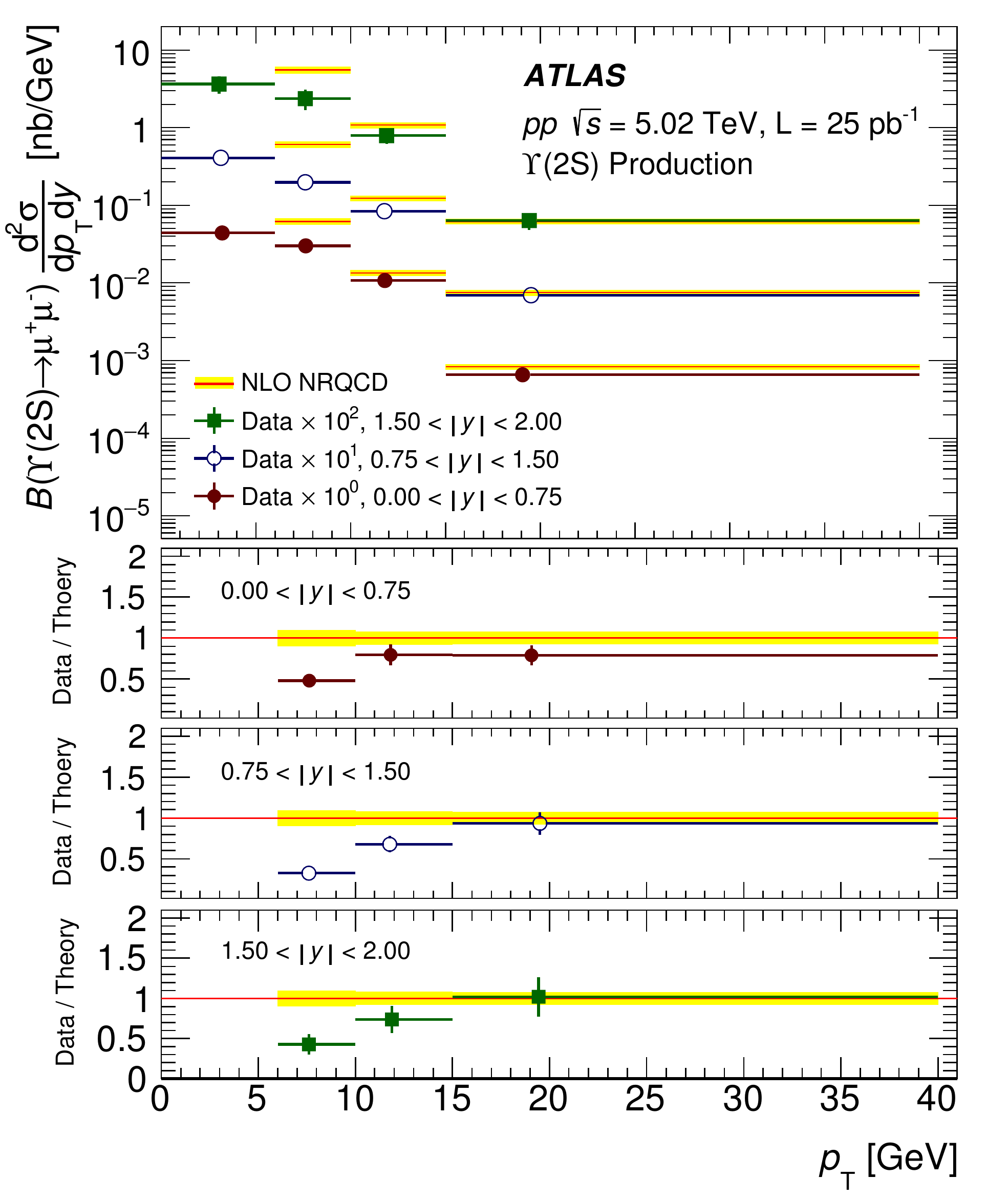}
  \includegraphics[width=\fourfigsize]{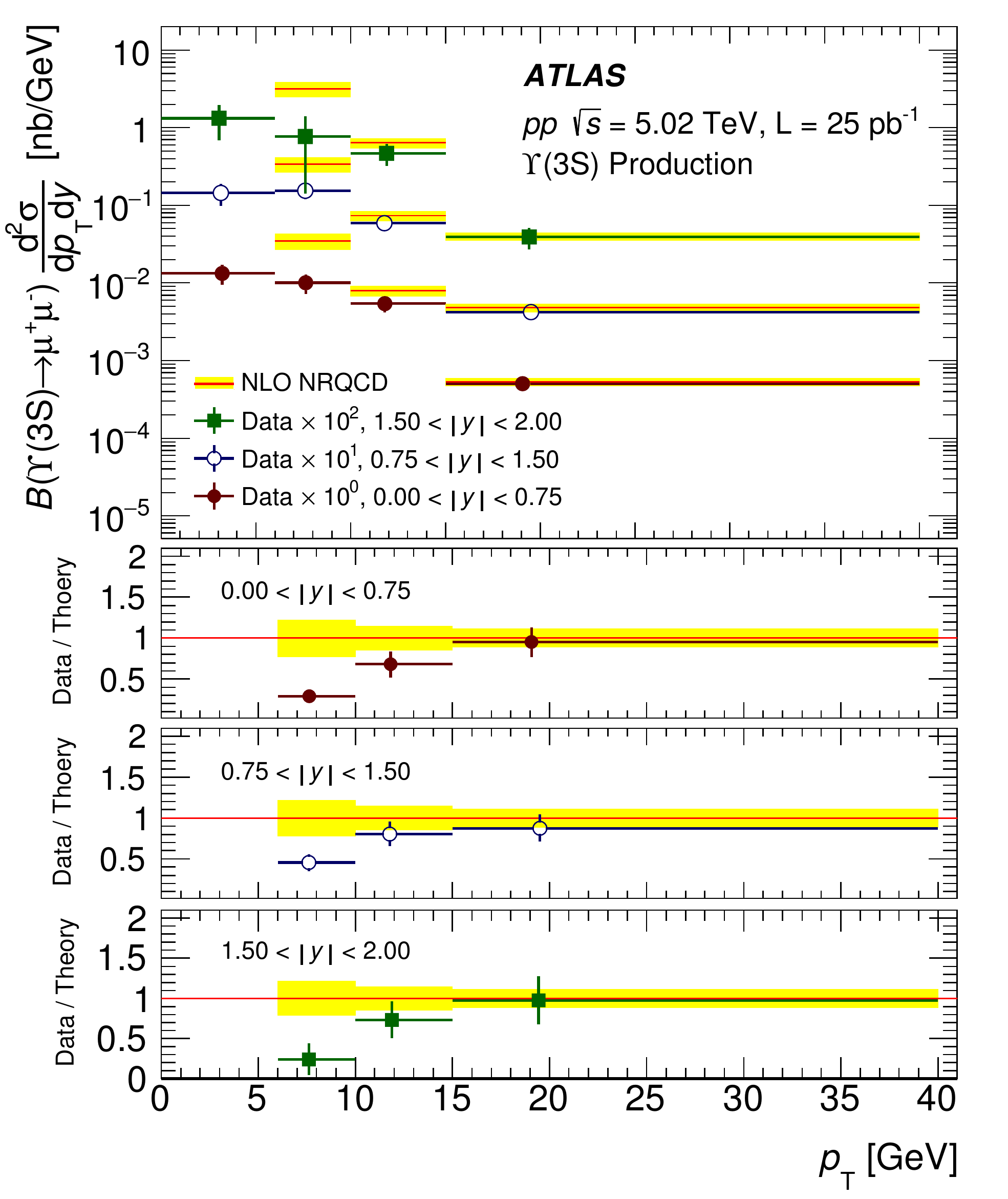}
  \caption{
  The differential production cross section times dimuon branching fraction of $\Upsii$ (left) and $\Upsiii$ (right)
  as a function of transverse momentum $\pt$ for three intervals of rapidity $y$ in $pp$ collisions at $5.02~\TeV$.
  For each increasing rapidity slice, an additional scaling factor of 10 is applied to the plotted points for visual clarity.
  The horizontal position of each data point indicates the mean of the weighted $\pt$ distribution.
  The horizontal bars represent the range of $\pt$ for the bin,
  and the vertical error bars correspond to the combined statistical and systematic uncertainties.
  The NRQCD theory predictions (see text) are also shown,
  and the error bands in the prediction correspond to the combined scale, quark mass and LDMEs uncertainties.
  }
  \label{fig:Upsi_Xsec_pp}
  \end{center}
\end{figure}

The measured prompt $\Jpsi$ and $\psip$ cross sections in $pp$ collisions at $\ppenergy$
are shown in Figure~\ref{fig:NLO_pp} in \pt and rapidity intervals, compared with non-relativistic QCD (NRQCD) predictions.
The theory predictions are based on the long-distance matrix elements
(LDMEs) from Refs.~\cite{Ma:2010yw,Shao:2014yta}, 
with uncertainties originating from the choice of scale, charm quark mass
and LDMEs (see Refs.~\cite{Ma:2010yw,Shao:2014yta} for more details).
Figures~\ref{fig:Upsi1S_Xsec_pp} and~\ref{fig:Upsi_Xsec_pp} show the production cross section of $\Upsins$ in $pp$ collisions 
compared to similar NRQCD model calculations~\cite{Han:2014kxa}.
As stated in Ref.~\cite{Han:2014kxa}, 
the LDMEs for bottomonium production are only extracted from fitting experiment data at $\pt > 15~\GeV$. 
At lower $\pt$, there might be non-perturbative effects which break the NRQCD factorization and perturbation expansion. 
As a consequence of its construction, the bottomonium NRQCD model gives a relatively good description of the measured $\Upsins$ production cross section at $\pt > 15~\GeV$, 
while overestimates the production cross section at lower $\pt$.

\begin{figure}[!htb]
  \begin{center}
  \includegraphics[width=\fourfigsize]{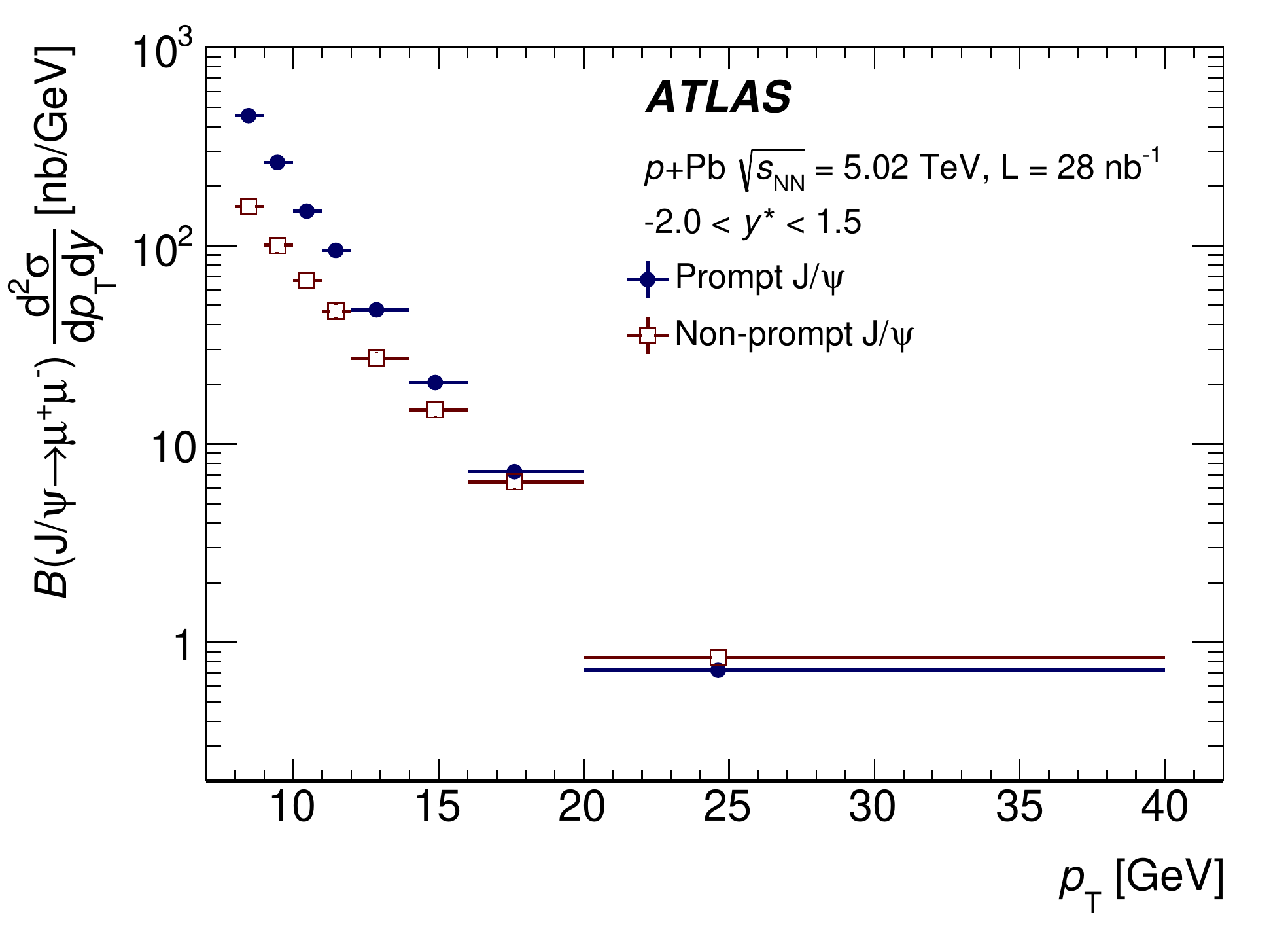}
  \includegraphics[width=\fourfigsize]{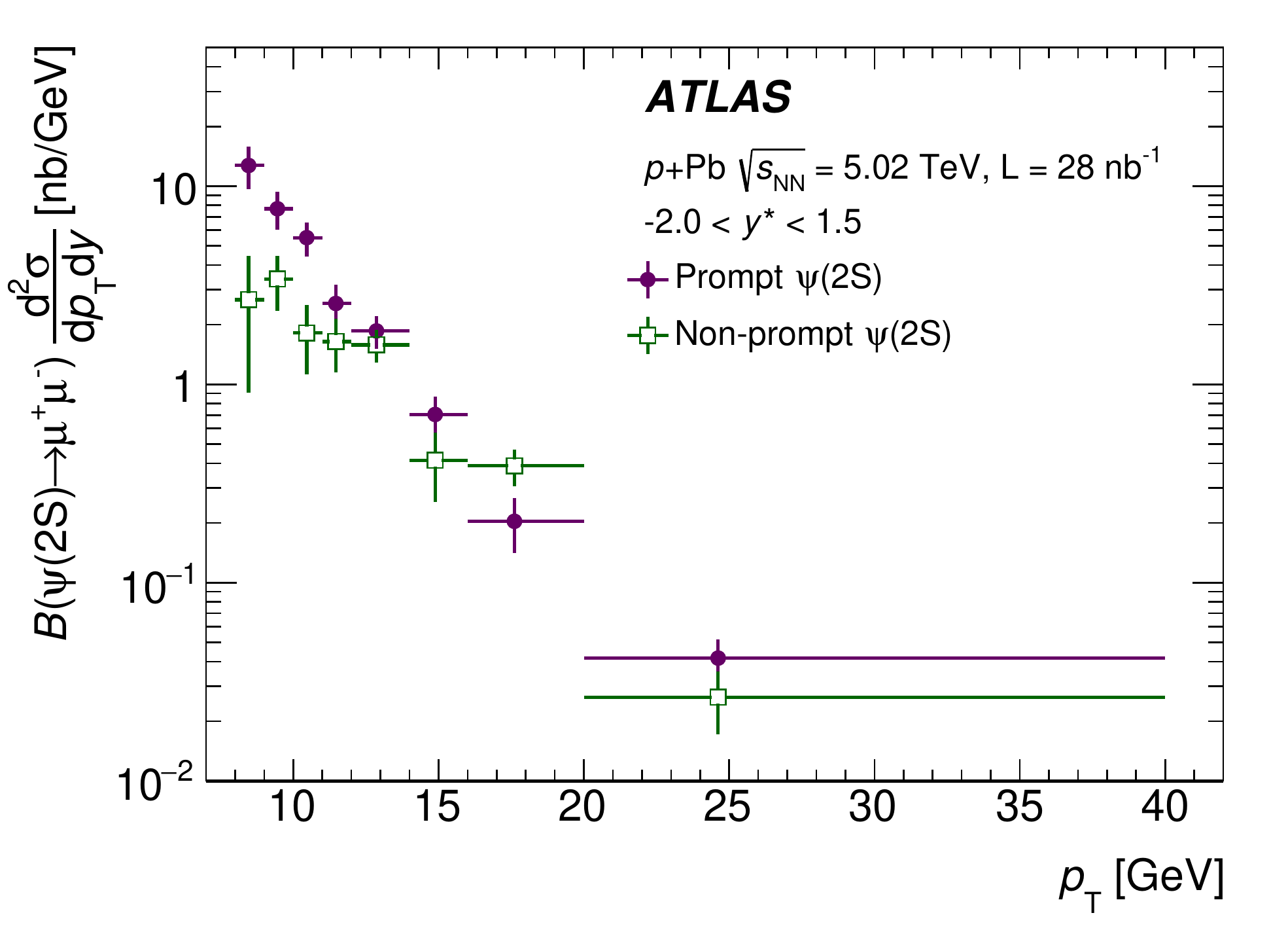}\\
  \caption{
  The differential cross section times dimuon branching fraction of prompt and non-prompt $\Jpsi$ (left) and $\psip$ (right)
  as a function of transverse momentum $\pt$ in \pPb\ collisions at $\energy$.
  The horizontal position of each data point indicates the mean of the weighted $\pt$ distribution.
  The horizontal bars represent the range of $\pt$ for the bin,
  and the vertical error bars correspond to the combined statistical and systematic uncertainties.
  }
  \label{fig:Jpsi_Xsec_pPb_dpT}
  \end{center}
\end{figure}

\begin{figure}[!htb]
  \begin{center}
  \includegraphics[width=\fourfigsize]{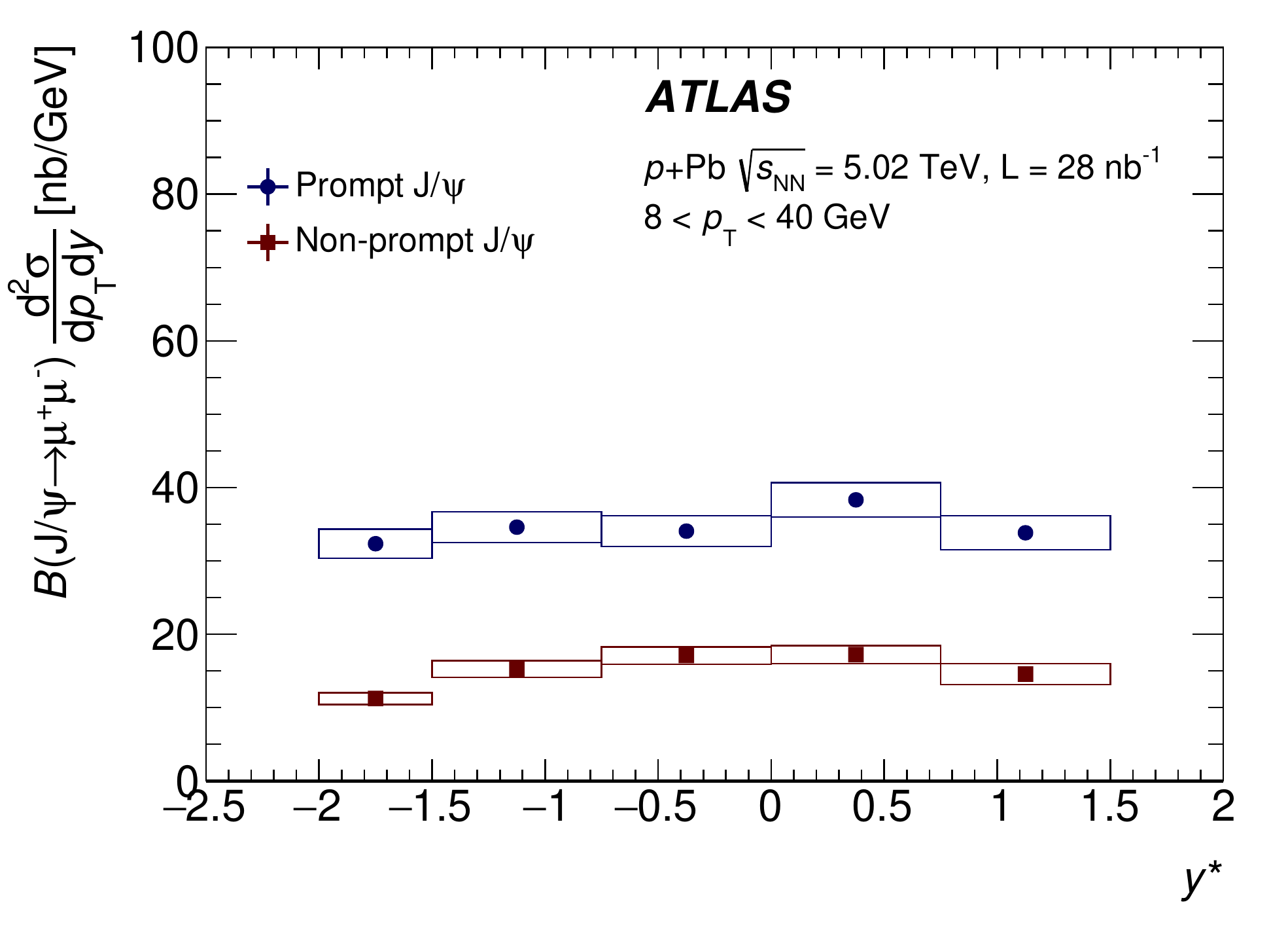}
  \includegraphics[width=\fourfigsize]{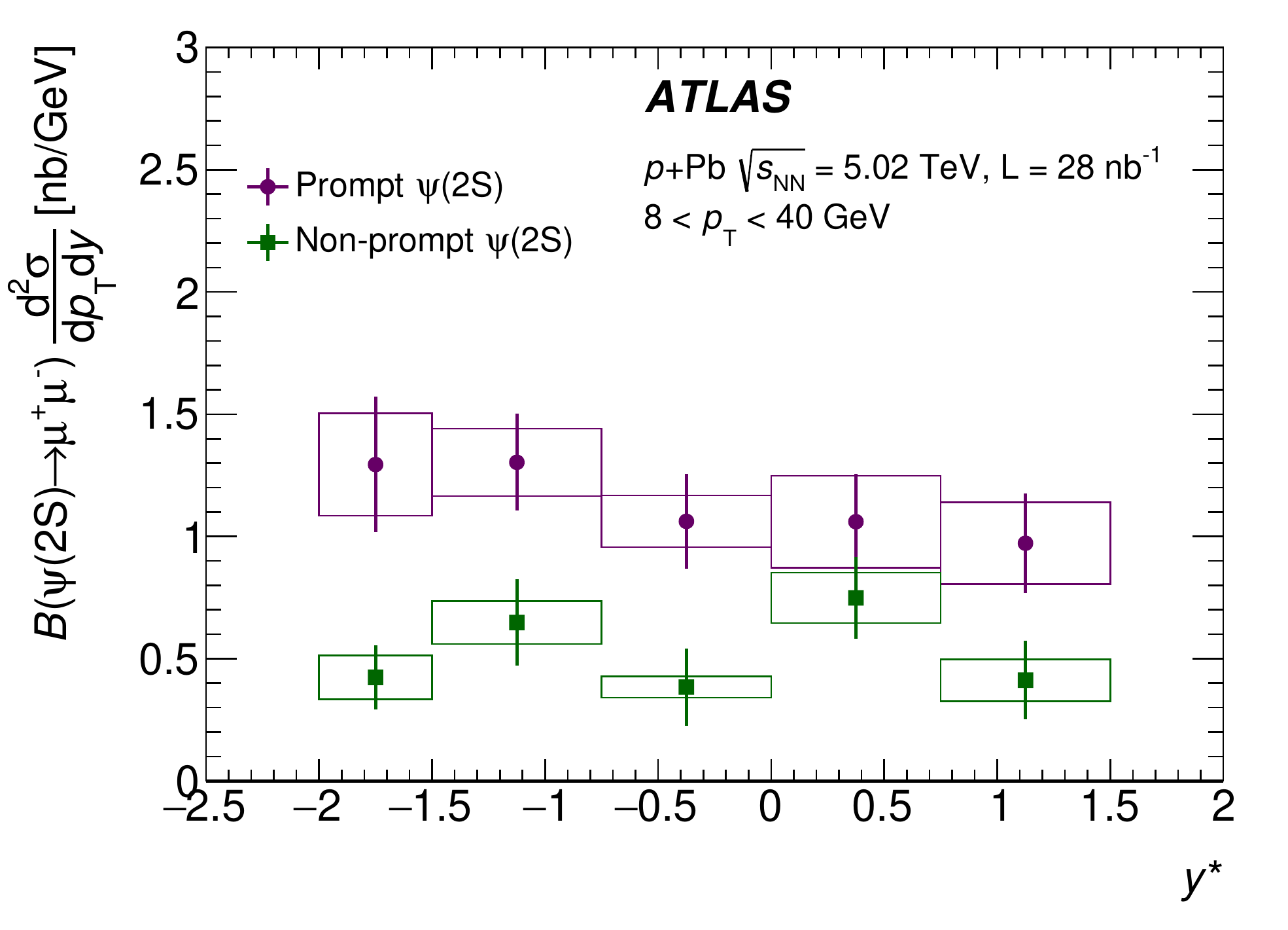}\\
  \caption{
  The differential cross section times dimuon branching fraction of prompt and non-prompt $\Jpsi$ (left) and $\psip$ (right)
  as a function of centre-of-mass rapidity $y^*$ in \pPb\ collisions at $\energy$.
  The horizontal position of each data point indicates the mean of the weighted $y^*$ distribution.
  The vertical error bars correspond to statistical uncertainties.
  The vertical sizes of coloured boxes around the data points represent the systematic uncertainties,
  and the horizontal sizes of coloured boxes represent the range of $y^*$ for the bin.
  }
  \label{fig:Jpsi_Xsec_pPb_dRap}
  \end{center}
\end{figure}

\begin{figure}[!htb]
  \begin{center}
  \includegraphics[width=\fourfigsize]{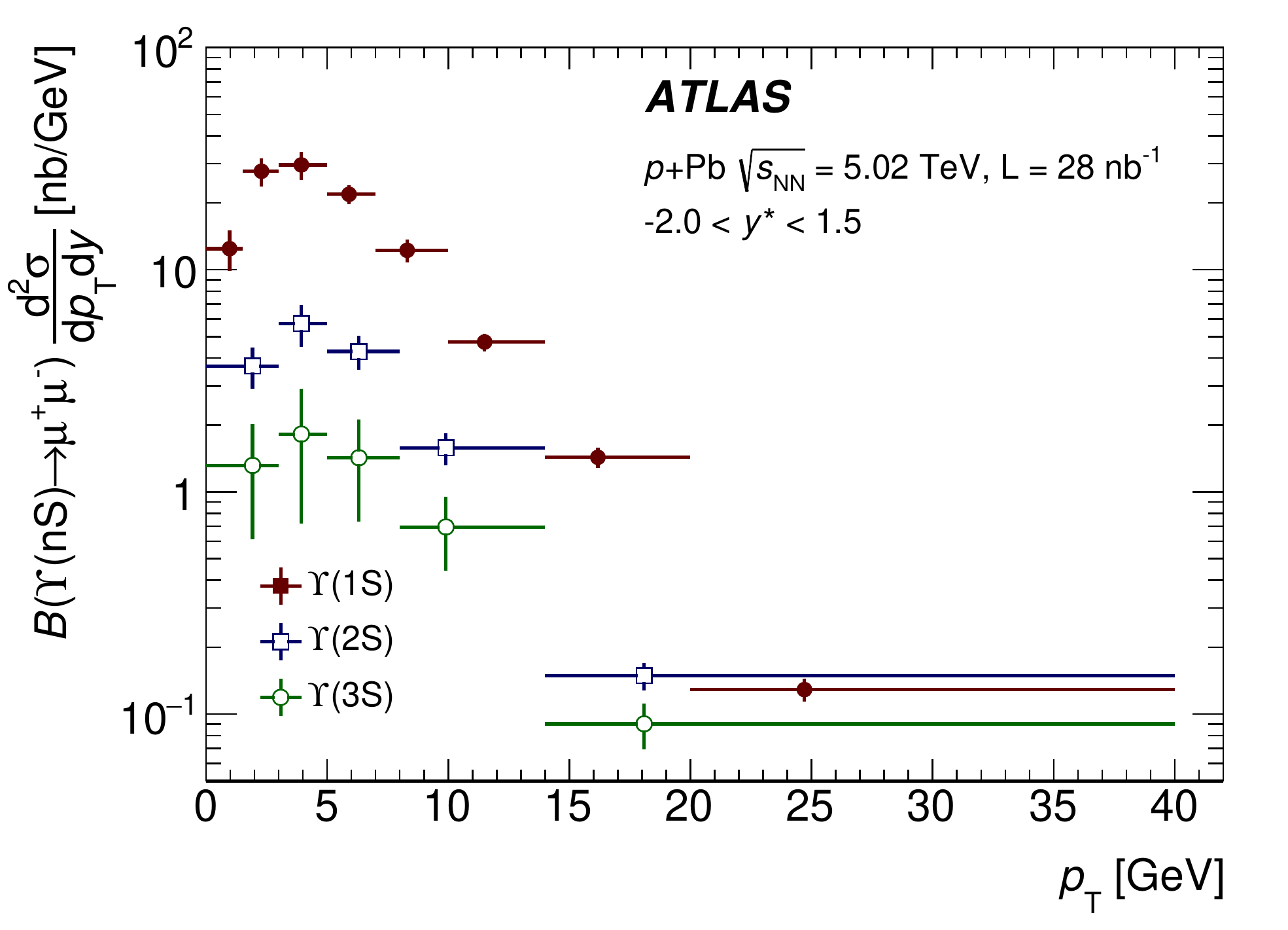}
  \includegraphics[width=\fourfigsize]{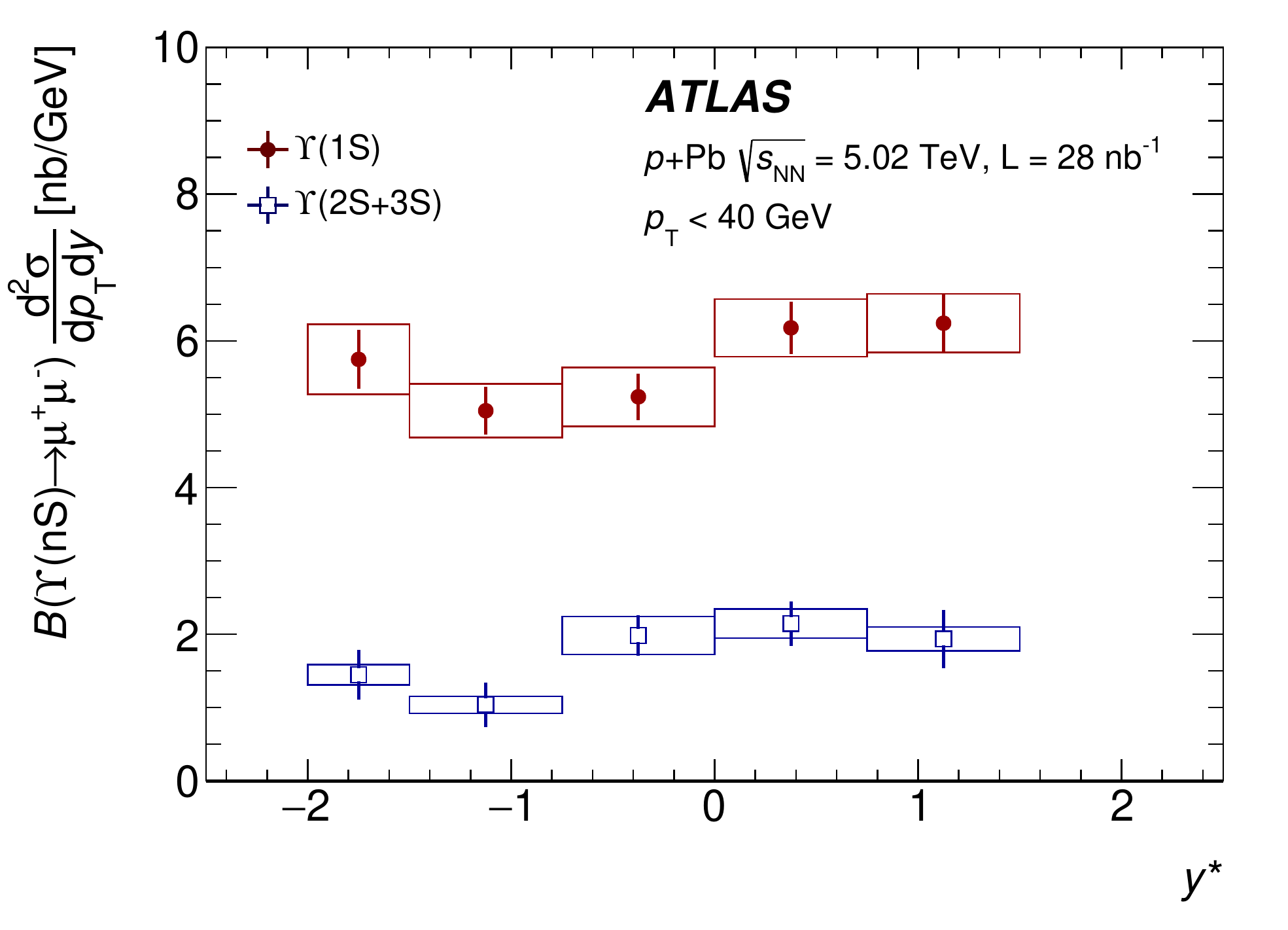}\\
  \caption{
  Left: the differential cross section times dimuon branching fraction of $\Upsins$
  as a function of transverse momentum $\pt$ in \pPb\ collisions at $\energy$.
  The horizontal position of each data point indicates the mean of the weighted $\pt$ distribution.
  The horizontal bars represent the range of $\pt$ for the bin,
  and the vertical error bars correspond to the combined statistical and systematic uncertainties.
  Right: the differential cross section times dimuon branching fraction of $\Upsins$
  as a function of centre-of-mass rapidity $y^*$ in \pPb\ collisions.
  The horizontal position of each data point indicates the mean of the weighted $y^*$ distribution.
  The vertical error bars correspond to statistical uncertainties.
  The vertical sizes of coloured boxes around the data points represent the systematic uncertainties,
  and the horizontal sizes of coloured boxes represent the range of $y^*$ for the bin.
  }
  \label{fig:Upsi_Xsec_pPb}
  \end{center}
\end{figure}

The results for prompt and non-prompt production cross sections of $\jpsi$ and $\psip$ in \pPb\ collisions at $\energy$
are shown in intervals of $\pt$ in Figure~\ref{fig:Jpsi_Xsec_pPb_dpT}.
The results for prompt and non-prompt production cross sections of $\jpsi$ and $\psip$ in \pPb\ collisions in intervals of $y^*$
are shown in Figure~\ref{fig:Jpsi_Xsec_pPb_dRap}.
Compared to the previous ATLAS measurement~\cite{HION-2013-07}, 
improved muon trigger corrections which are smaller by $6\%$ in central value and $4\%$ in uncertainty
and a more comprehensive fit model involving a wider mass range 
are used in the $\Jpsi$ cross-section measurements.
The measured $\Jpsi$ cross sections are consistent with previous results within uncertainties.
The measured differential production cross section of $\Upsins$ in \pPb collisions is shown in Figure~\ref{fig:Upsi_Xsec_pPb}.
Due to difficulties in separating $\Upsii$ and $\Upsiii$ at forward and backward $y^*$ intervals in \pPb collisions, 
they are combined to obtain stable rapidity dependence of the production cross section.

\subsection{Nuclear modification factor}

\begin{figure}[!htb]
  \begin{center}  
  \includegraphics[width=\fourfigsize]{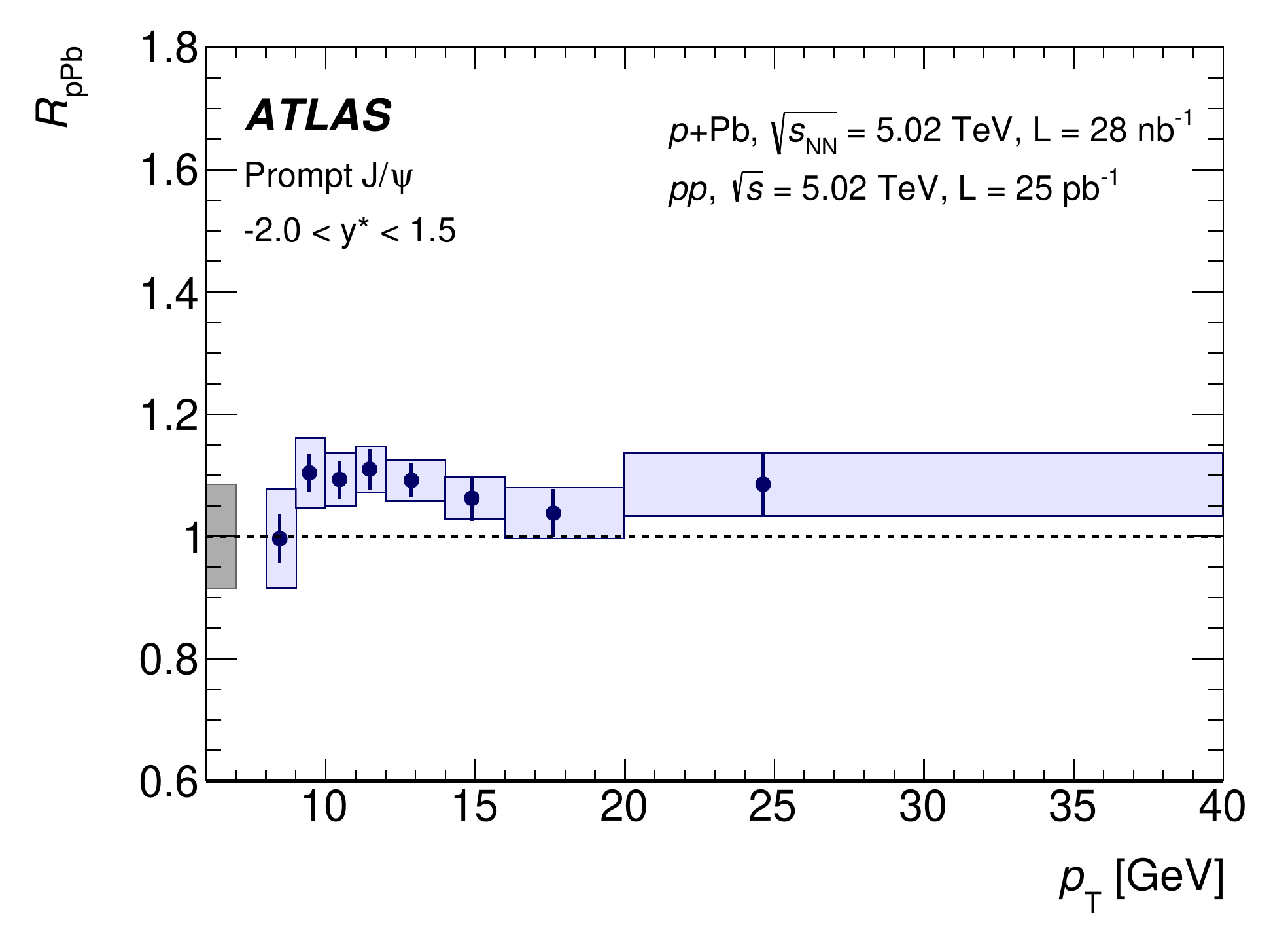}
  \includegraphics[width=\fourfigsize]{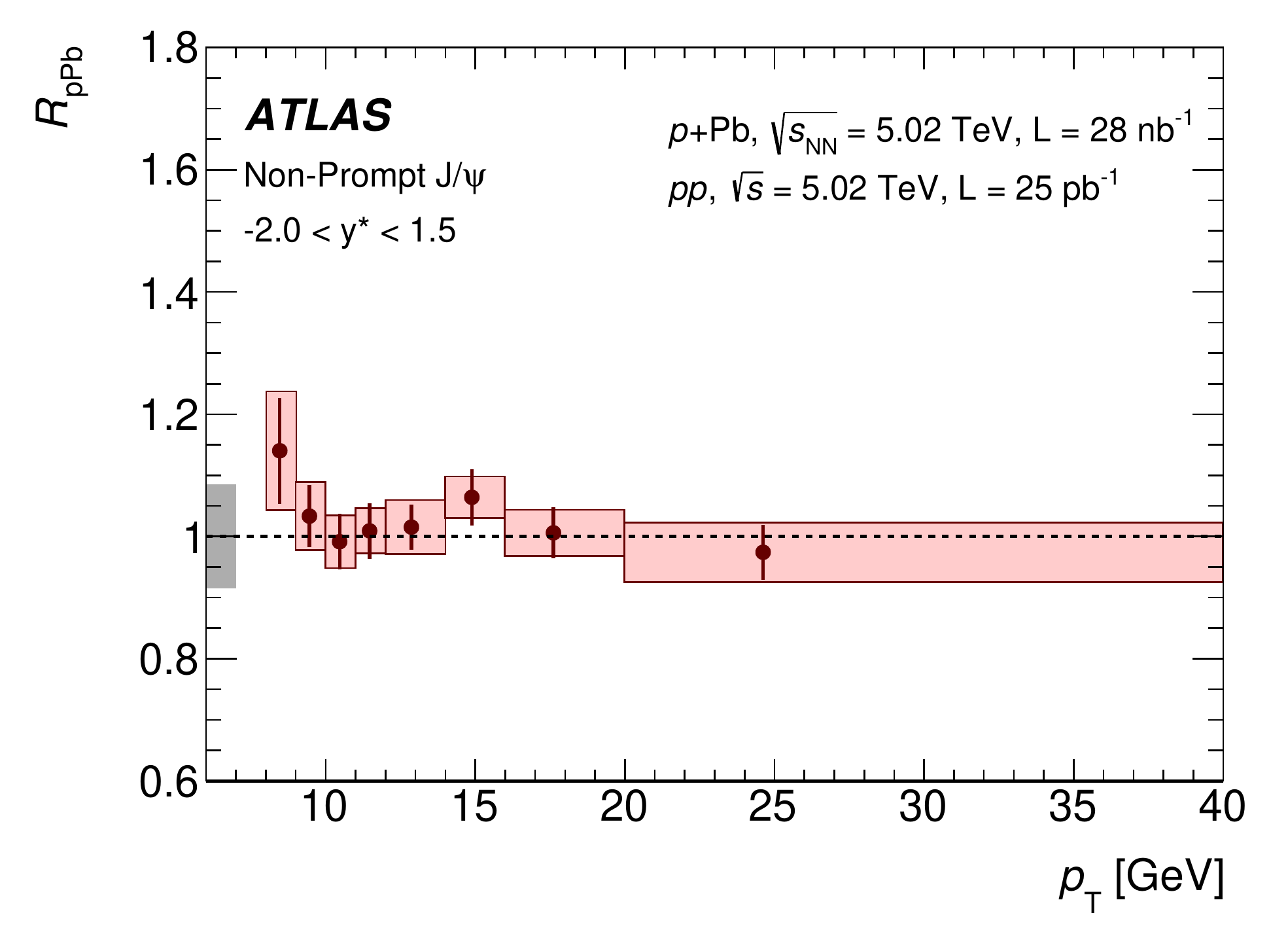}\\
  \caption{The nuclear modification factor, $\rpb$, as a function of transverse momentum $\pt$ for prompt $\Jpsi$ (left) and non-prompt $\Jpsi$ (right).
  The horizontal position of each data point indicates the mean of the weighted $\pt$ distribution.
  The vertical error bars correspond to the statistical uncertainties.
  The vertical sizes of coloured boxes around the data points represent the uncorrelated systematic uncertainties,
  and the horizontal sizes of coloured boxes represent the $\pt$ bin sizes.
  The vertical sizes of the leftmost grey boxes around $\rpb = 1$ represent the correlated systematic uncertainty.
  }
  \label{fig:jpsi_RpPb}
  \end{center}
\end{figure}

\begin{figure}[!htb]
  \begin{center}  
  \includegraphics[width=\fourfigsize]{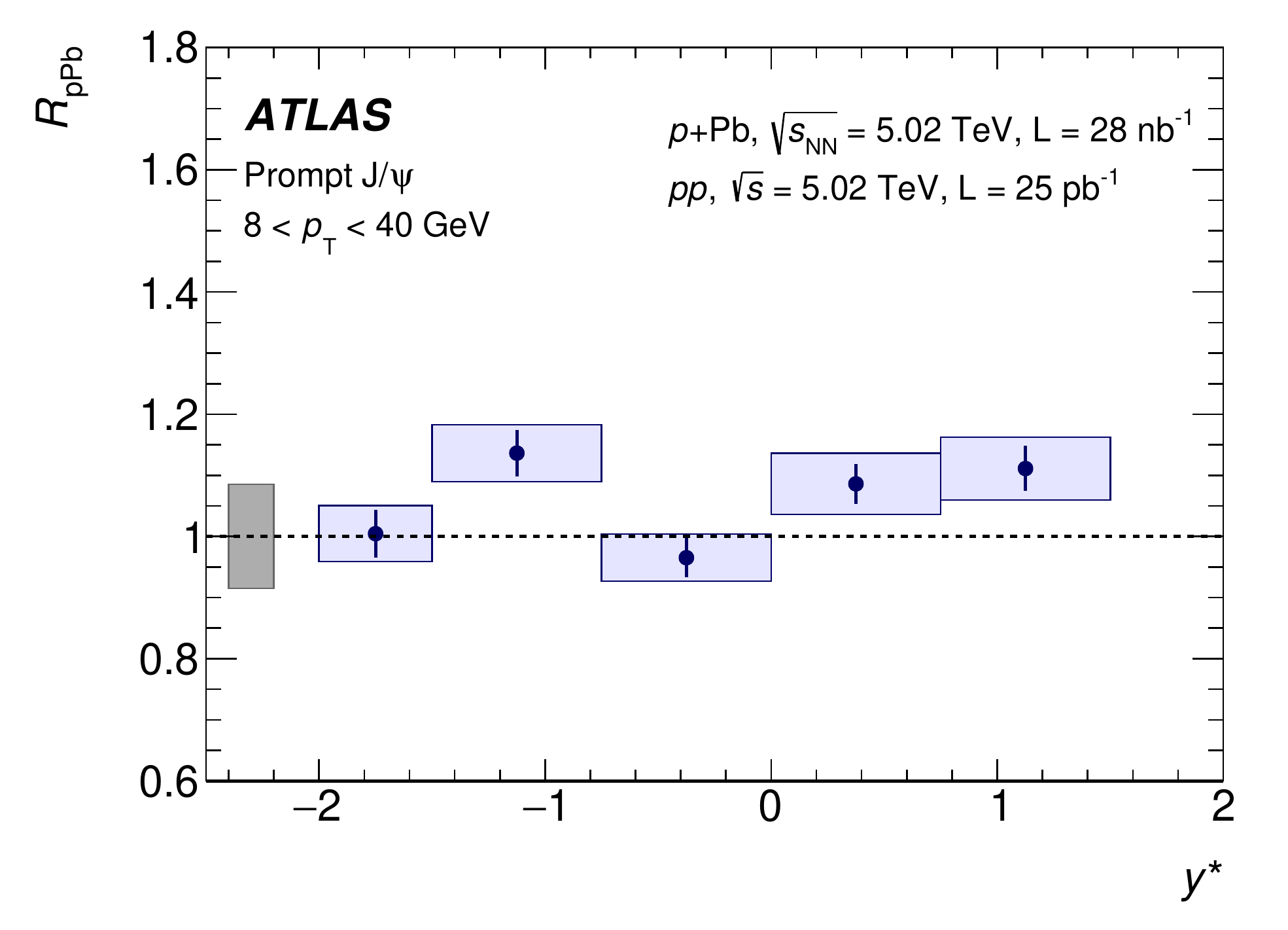}
  \includegraphics[width=\fourfigsize]{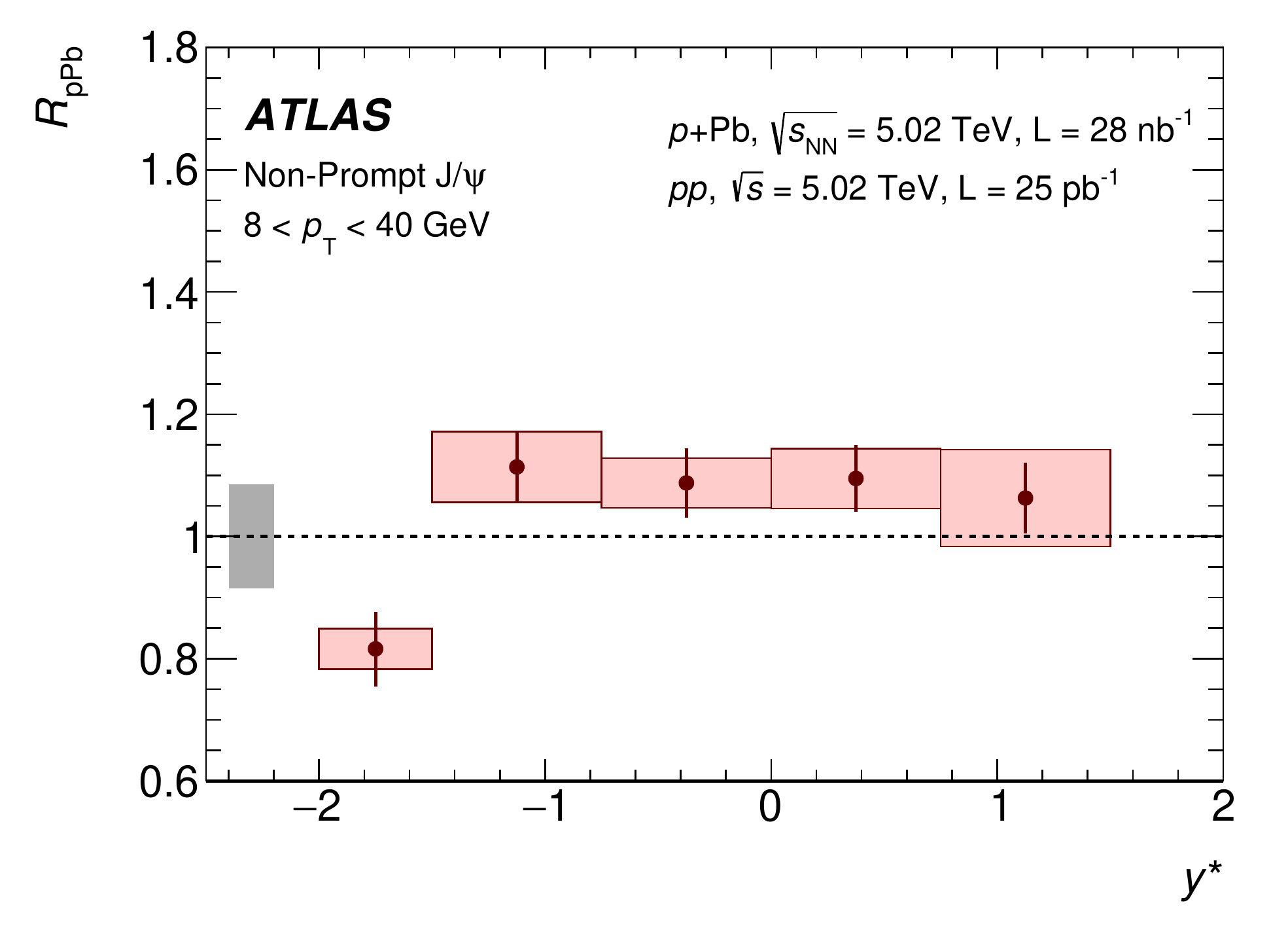}\\
  \caption{The nuclear modification factor, $\RpPb$, as a function of centre-of-mass rapidity $y^*$ for prompt $\Jpsi$ (left) and non-prompt $\Jpsi$ (right).
  The horizontal position of each data point indicates the mean of the weighted $y^*$ distribution.
  The vertical error bars correspond to the statistical uncertainties.
  The vertical sizes of coloured boxes around the data points represent the uncorrelated systematic uncertainties,
  and the horizontal sizes of coloured boxes represent the $y^*$ bin sizes.
  The vertical sizes of the leftmost grey boxes around $\rpb = 1$ represent the correlated systematic uncertainty.
  }
  \label{fig:jpsi_RpPb_y}
  \end{center}
\end{figure}

\begin{figure}[!htb]
  \begin{center}
  
  \includegraphics[width=\fourfigsize]{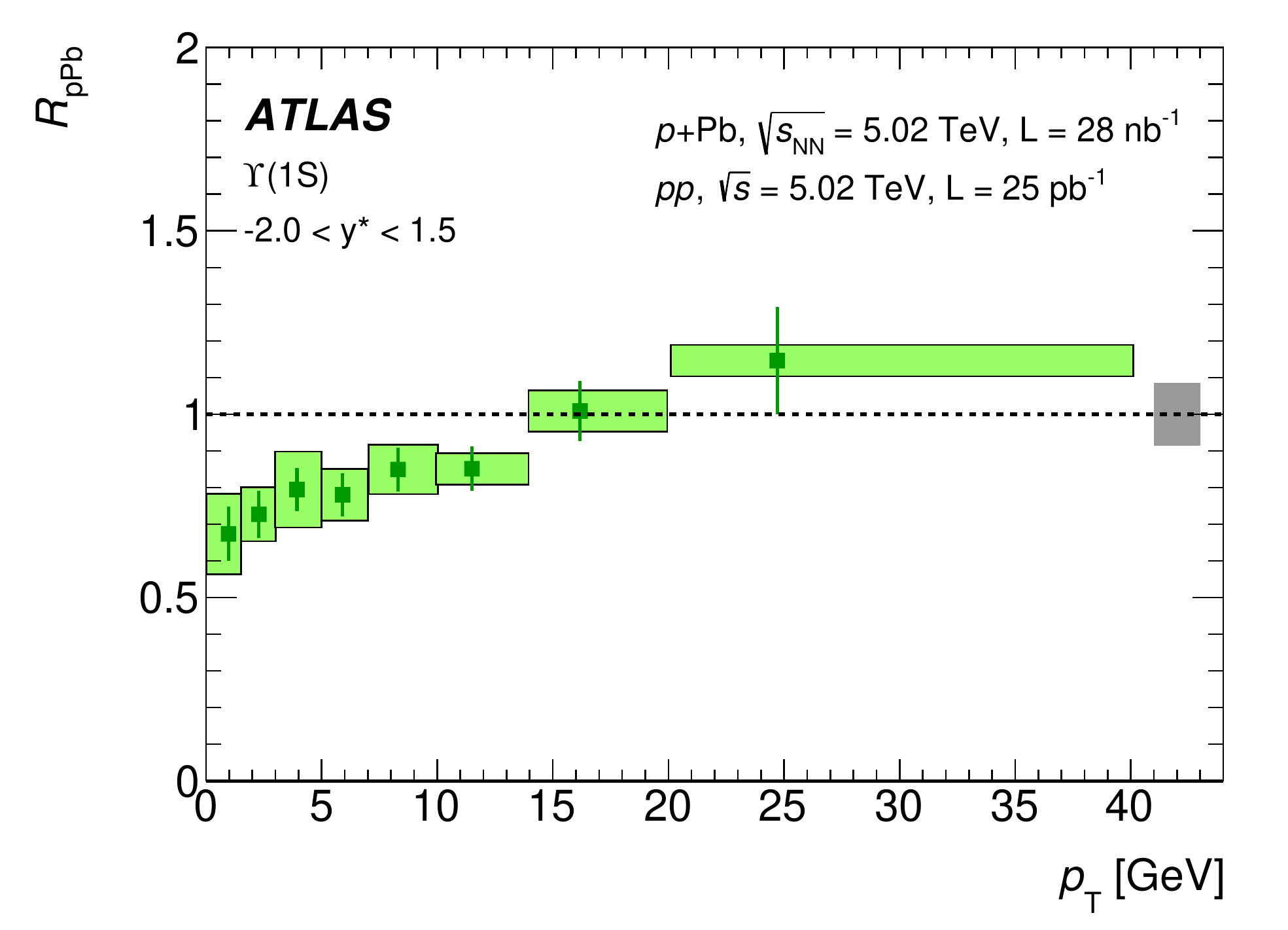}
  \includegraphics[width=\fourfigsize]{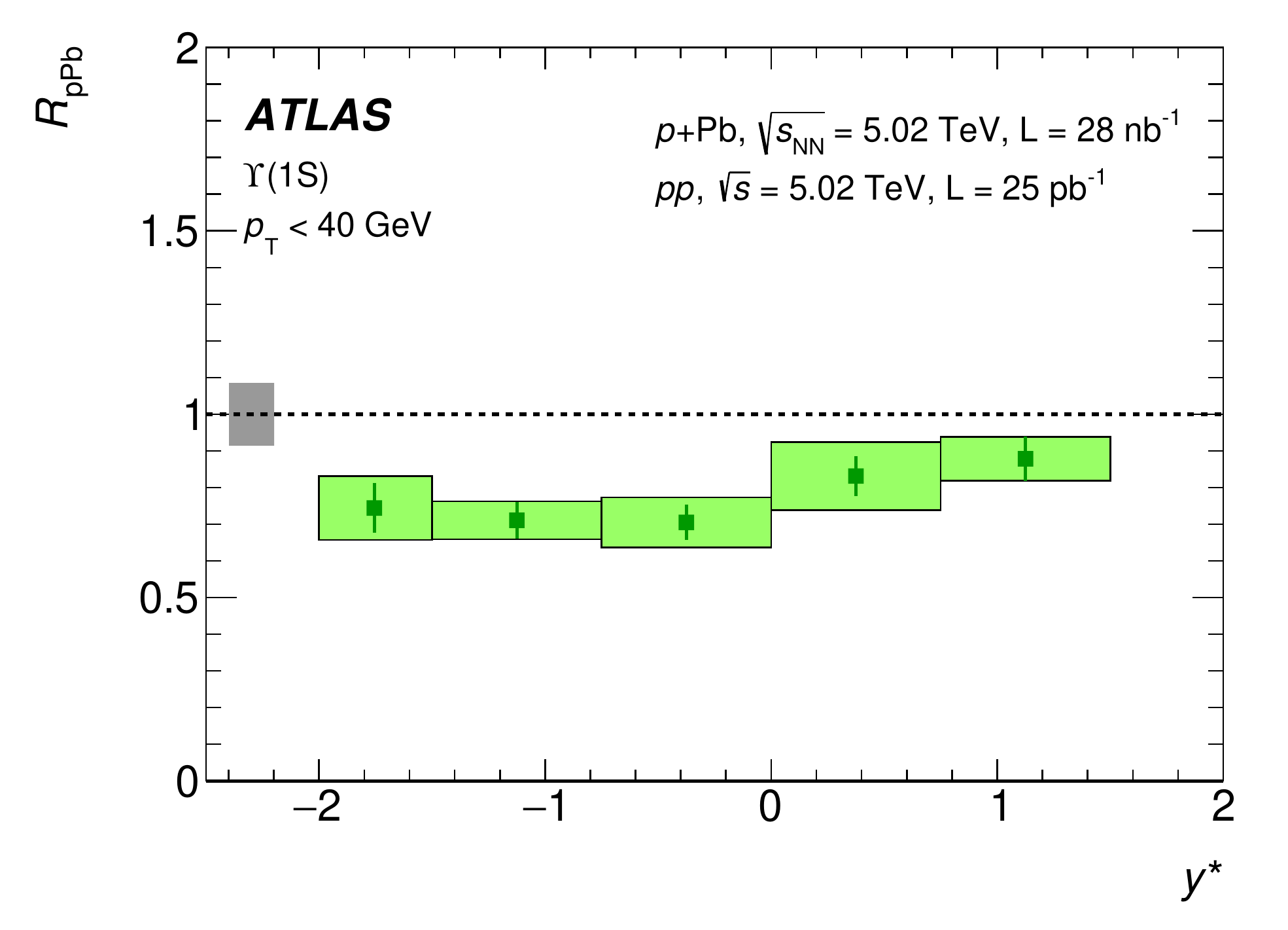}
  \caption{The nuclear modification factor, $\RpPb$, as a function of transverse momentum $\pt$ (left) and centre-of-mass rapidity $y^*$ (right) for $\Upsi$.
  The horizontal position of each data point indicates the mean of the weighted $\pt$ or $y^*$ distribution.
  The vertical error bars correspond to the statistical uncertainties.
  The vertical sizes of coloured boxes around the data points represent the uncorrelated systematic uncertainties,
  and the horizontal sizes of coloured boxes represent the bin sizes.
  The vertical size of the rightmost (left) and leftmost (right) grey boxes around $\rpb = 1$ represent the correlated systematic uncertainty.
  }
  \label{fig:Upsi_RpPb}
  \end{center}
\end{figure}

The $\pt$ dependence of $\RpPb$ for the prompt and non-prompt $\jpsi$ is shown in Figure~\ref{fig:jpsi_RpPb}.
Taking into account the correlated and uncorrelated uncertainties,
both the prompt and non-prompt $\jpsi$ $\RpPb$ are consistent with unity across the $\pt$ range from 8 to $40~\GeV$.
The rapidity dependence of prompt and non-prompt $\jpsi$ $\RpPb$ is shown in Figure~\ref{fig:jpsi_RpPb_y}.
No significant rapidity dependence is observed.
The $\pt$ and rapidity dependence of $\Upsi$ $\RpPb$ is shown in Figure~\ref{fig:Upsi_RpPb}.
The $\Upsi$ production in \pPb collisions is found to be suppressed compared to \pp collisions at low $\pt$ ($\pt< 15~\GeV$),
and increases with \pt.
Low \pt $\Upsi$ can probe smaller Bjorken-x region compared to $\jpsi$ measured in $8 < \pt < 40~\GeV$~\cite{Albacete:2013ei},
so the observed suppression of $\Upsi$ production at low \pt may come from the reduction of hard-scattering cross sections 
due to stronger nPDF shadowing at smaller Bjorken-x.
No significant rapidity dependence is observed,
which qualitatively agrees with a prediction of weak rapidity dependence for central rapidities.

\begin{figure}[!htb]
  \begin{center}
  \includegraphics[width=\fourfigsize]{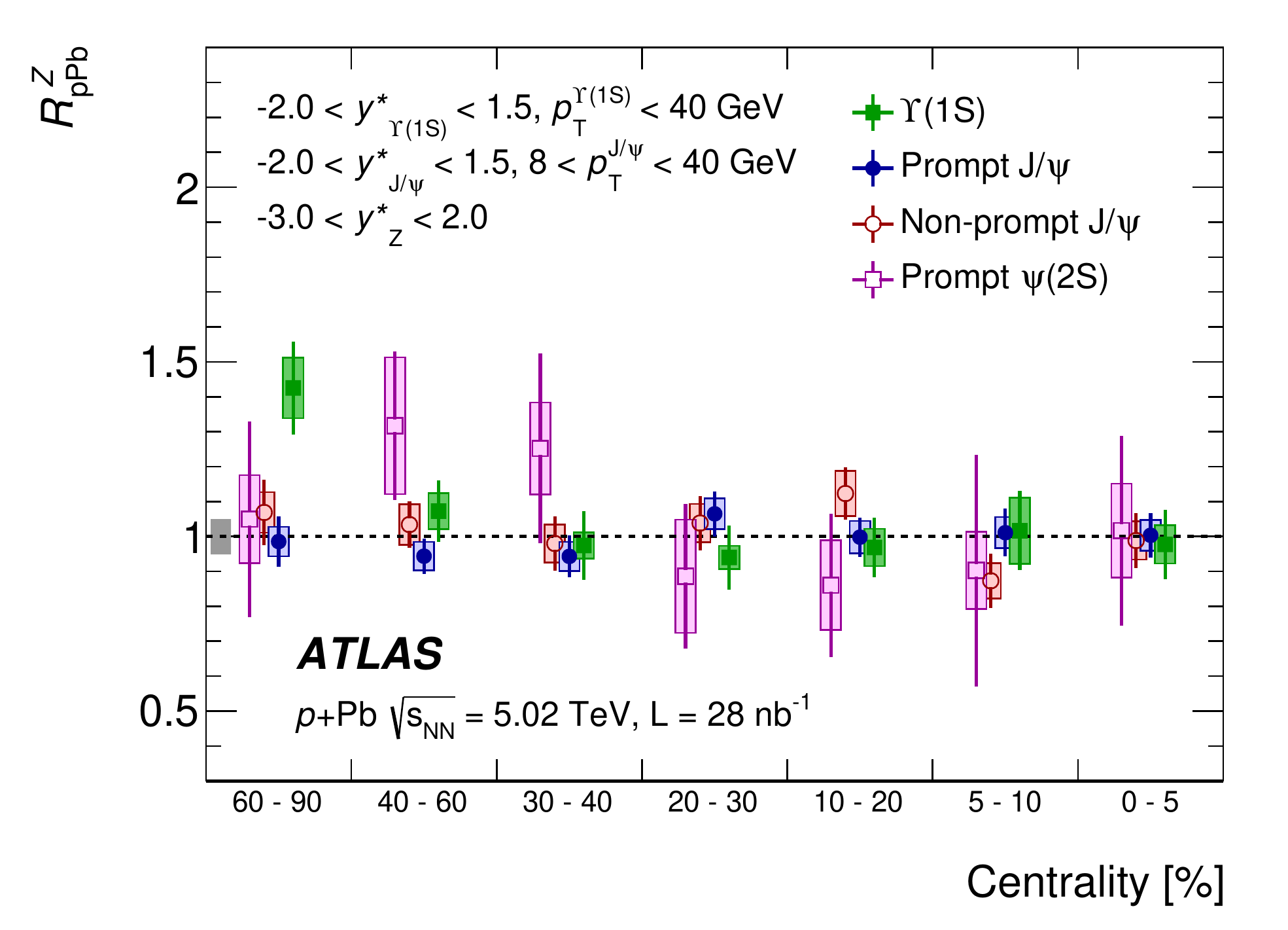}
  \caption{Prompt and non-prompt $\Jpsi$, prompt $\psip$ and $\Upsi$ to $Z$ boson yield ratio, $\rpb^Z$, as a function of event centrality in \pPb collisions.
  The ratio is normalised to the ratio integrated in the centrality range $0$--$90\%$.
  The error bars represent statistical uncertainties.
  The vertical sizes of coloured boxes around the data points represent the uncorrelated systematic uncertainties,
  and the vertical size of the leftmost grey box around $\rpb^Z = 1$ represents the correlated systematic uncertainty.
  }
  \label{fig:RatioOverZ}
  \end{center}
\end{figure}

The $Z$ boson does not interact with the nuclear medium via the strong interaction, 
so it is considered a good reference process in \pPb collisions for studying the centrality dependence of quarkonium production in a model-independent way.
The quarkonium yield is compared to the $Z$ boson yield from Ref.~\cite{HION-2013-09} in intervals of centrality.
The ratio of quarkonium to the $Z$ boson yield is defined as:

\begin{equation*}
	\rpb^{Z} (\mathcal{O}(n\text{S})) = \dfrac{ N_{\mathcal{O}(n\text{S})}^{\text{cent}} / N_Z^{\text{cent}} }{ N_{\mathcal{O}(n\text{S}) }^{0-90\%} / N_Z^{0-90\%} }
\end{equation*}
where $N_{\mathcal{O}(n\text{S})}^{\mathrm{cent}}$ ($N_Z^{\text{cent}}$) is the corrected quarkonium ($Z$ boson) yield for one centrality class.
The resulting $\rpb^{Z} (\mathcal{O}(n\text{S}))$ is shown in Figure~\ref{fig:RatioOverZ} for the different quarkonium states in intervals of centrality.
In each centrality interval, $\rpb^{Z} (\mathcal{O}(n\text{S}))$ is normalised to the ratio integrated in the centrality range $0$--$90\%$
such that the normalised yield ratio is independent of production cross sections of the different quarkonium states.
The prompt and non-prompt $\jpsi$ are found to behave in a way very similar to the $Z$ boson.
The $Z$ boson production is found to scale with the number of binary collisions in \pPb collisions after applying the centrality bias correction factor~\cite{HION-2013-09, Perepelitsa:2014yta}.
The centrality bias correction factor proposed in Ref.~\cite{Perepelitsa:2014yta} does not depend on the physics process, 
so the measured $\rpb^{Z} (\Jpsi)$ suggests that centrality-bias-corrected $\jpsi$ production also scales with the number of binary collisions.
The measured $\rpb^{Z} (\Upsi)$ is consistent with being a constant except for the measurement in the most peripheral ($60$--$90\%$) \pPb\ collisions,
which is about two to three standard deviations away from the value observed in more central collisions.
The current precision of $\rpb^{Z} (\psip)$ does not allow conclusions to be drawn about the 
centrality dependence of prompt $\psip$ production with respect to $Z$ bosons.

\begin{figure}[!htb]
  \begin{center}
  \includegraphics[width=\fourfigsize]{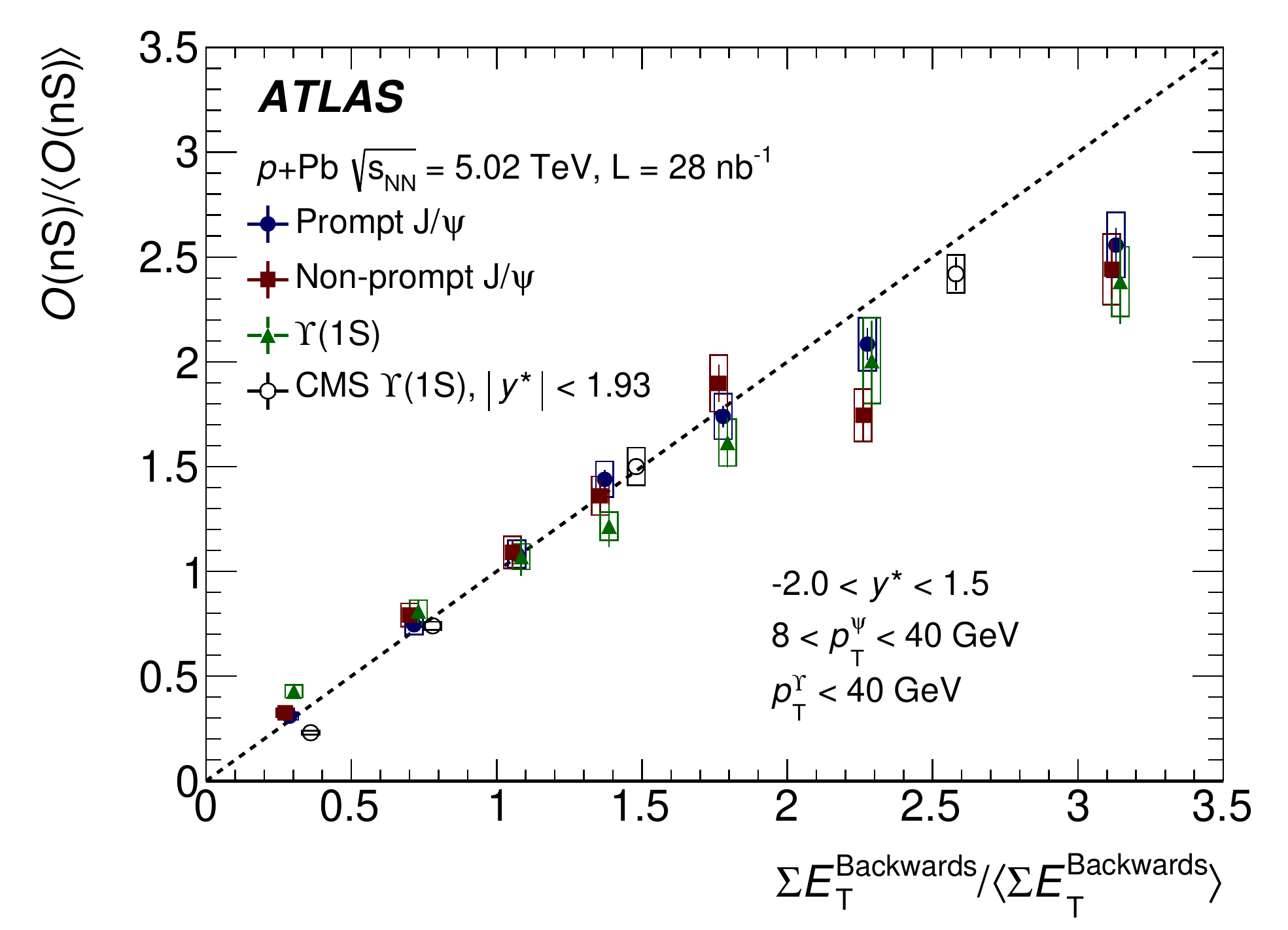}
  \caption{
  Self-normalised yield for prompt $\jpsi$, non-prompt $\jpsi$ and $\Upsi$,
  compared to $\Upsi$ self-normalised yield ratio measured by CMS~\cite{CMS-HIN-13-003}.
  The horizontal position of each data point represents the normalised mean value of the $\Sigma\ET^{\mathrm{Backwards}}$ distribution in minimum-bias data sample.
  The error bar represents statistical uncertainties,
  and the vertical size of box underneath the data point represents the systematic uncertainties.
  The dotted line is a linear function with a slope equal to unity.
  }
  \label{fig:SelfRatio}
  \end{center}
\end{figure}

Quarkonium self-normalised yields, ${ \mathcal{O}(n\text{S}) }/{\langle \mathcal{O}(n\text{S}) \rangle}$,
are defined as the per-event yields of quarkonium in each centrality class normalised by the yield in the $0$--$90\%$ centrality interval.
The correlation of quarkonium production with the underlying event
is traced by comparing the self-normalised quarkonium yields with the respective self-normalised event activity.
The event activity is characterised by the total transverse energy deposition in the backward FCal ($3.1 < |\eta| < 4.9$), 
$\Sigma\ET^{\mathrm{Backwards}}$, on the Pb-going side, 
and it is determined in a minimum-bias data sample as used in Ref.~\cite{HION-2012-15}.
The self-normalised quantities ${ \mathcal{O}(n\text{S}) }/{\langle \mathcal{O}(n\text{S}) \rangle}$ and ${\Sigma\ET^{\mathrm{Backwards}}}/{\langle\Sigma\ET^{\mathrm{Backwards}}\rangle}$ are defined as:

\begin{equation*}
        \frac{\mathcal{O}(n\text{S})}{\langle \mathcal{O}(n\text{S}) \rangle} \equiv \frac{N_{\mathcal{O}(n\text{S})}^{\mathrm{cent}}/N_{\mathrm{evt}}^{\mathrm{cent}} }{ N_{\mathcal{O}(n\text{S})}^{~0-90\%}/N_{\mathrm{evt}}^{~0-90\%} },\\
\label{eq:SR_y}
\end{equation*}

\begin{equation*}
        \frac{ \Sigma\ET^{\mathrm{Backwards}} }{\langle \Sigma\ET^{\mathrm{Backwards}} \rangle} = \frac{\langle \Sigma\ET^{\mathrm{Backwards}} \rangle_{~\mathrm{cent}}}{\langle \Sigma\ET^{\mathrm{Backwards}} \rangle_{~0-90\%}},
\label{eq:SR_x}
\end{equation*}
where $N_{\mathrm{evt}}^{\mathrm{cent}}$ is the number of events in the minimum-bias sample for one centrality class.
The measured self-normalised yields for prompt $\jpsi$, non-prompt $\jpsi$ and $\Upsi$
in \pPb\ collisions are shown in Figure~\ref{fig:SelfRatio}
in comparison with the same observable for $\Upsi$ in a previous CMS measurement~\cite{CMS-HIN-13-003}.
The event activity is determined in the range $4.0 < |\eta| < 5.2$ in CMS.
The $\Upsi$ self-normalised yields from ATLAS and CMS show a consistent trend.
In the events with the highest event activity, a two-standard-deviation departure from the linear trend is observed.
Since the same centrality dependence is found for ground-state quarkonium states and Z bosons as seen in Figure~\ref{fig:RatioOverZ},
the deviation at highest event activity may suggest that the $\Sigma\ET^{\mathrm{Backwards}}$ characterised event activity is not a robust scale parameter, 
but a more complicated geometry model is needed for instance as discussed in Ref.~\cite{HION-2013-09}.

\subsection{Double ratio}

\begin{figure}[!htb]
  \begin{center}
  \includegraphics[width=\fourfigsize]{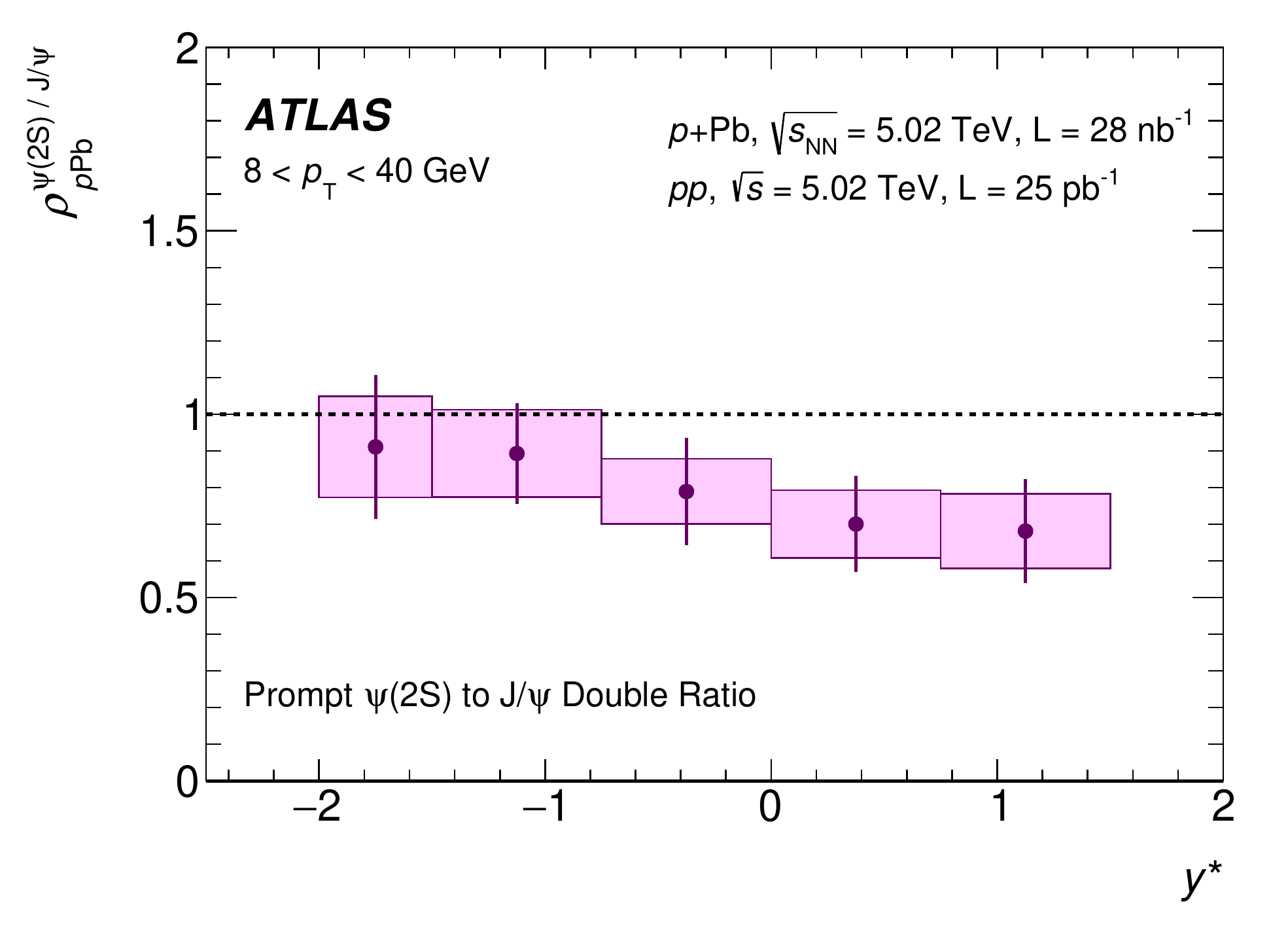}
  \caption{The prompt charmonium production double ratio, $\rhopbc$, as a function of the centre-of-mass rapidity, $y^*$.
  The vertical error bars correspond to the statistical uncertainties.
  The horizontal position of each data point indicates the mean of the weighted $y^*$ distribution.
  The vertical sizes of coloured boxes around the data points represent the uncorrelated systematic uncertainties,
  and horizontal sizes of the coloured boxes represent the bin sizes.
  }
  \label{fig:DR_EtRap}
  \end{center}
\end{figure}

\begin{figure}[!htb]
  \begin{center}
  \includegraphics[width=\fourfigsize]{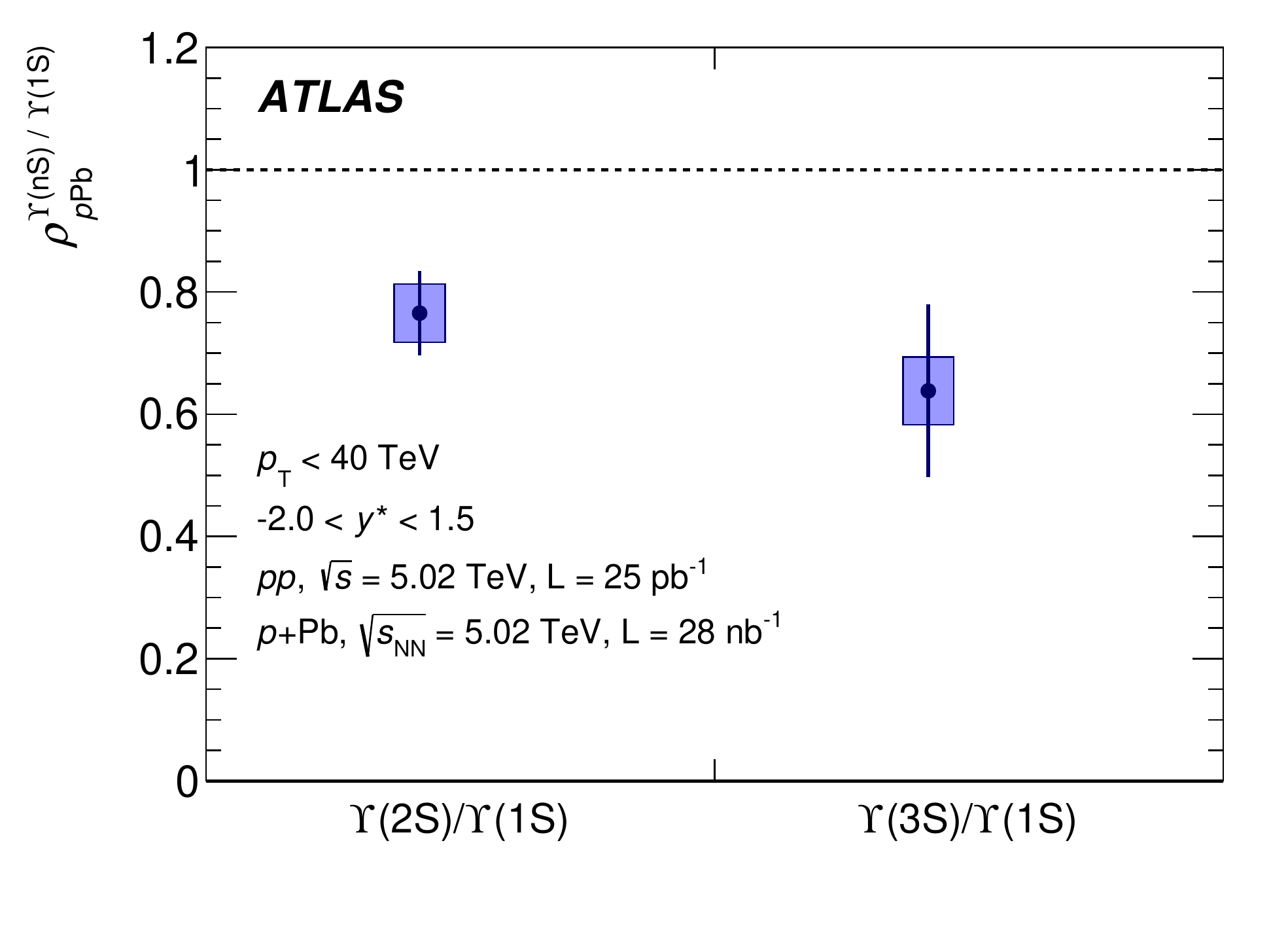}
  \caption{The bottomonium double ratio, \rhopbb , integrated in the whole measured $\pt$ and $y^*$ range.
  The vertical error bars correspond to the statistical uncertainties.
  The vertical sizes of boxes around the data points represent the systematic uncertainties.
  }
  \label{fig:DR_Upsi}
  \end{center}
\end{figure}

\begin{figure}[!htb]
  \begin{center}
  \includegraphics[width=\fourfigsize]{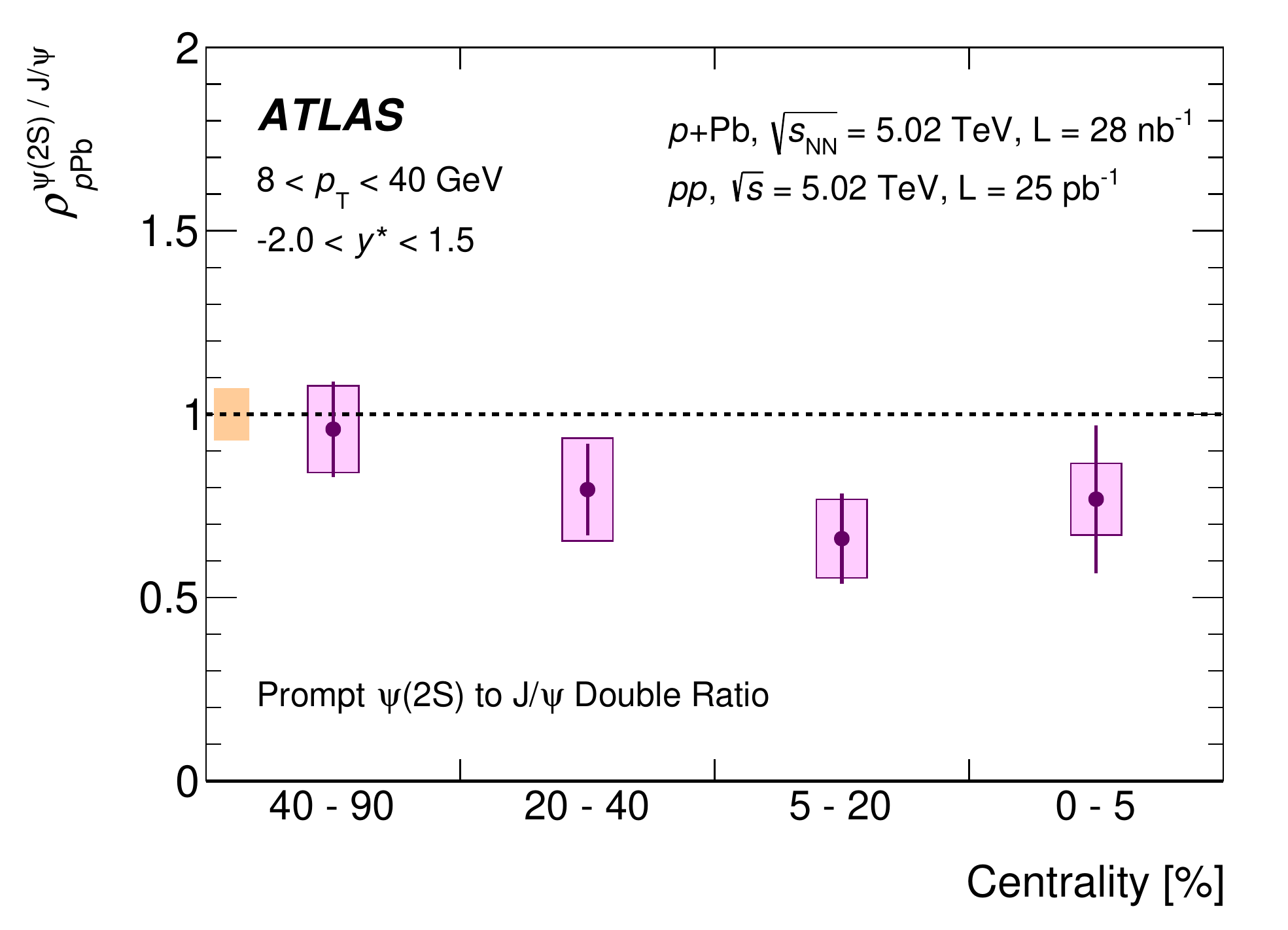}
  \includegraphics[width=\fourfigsize]{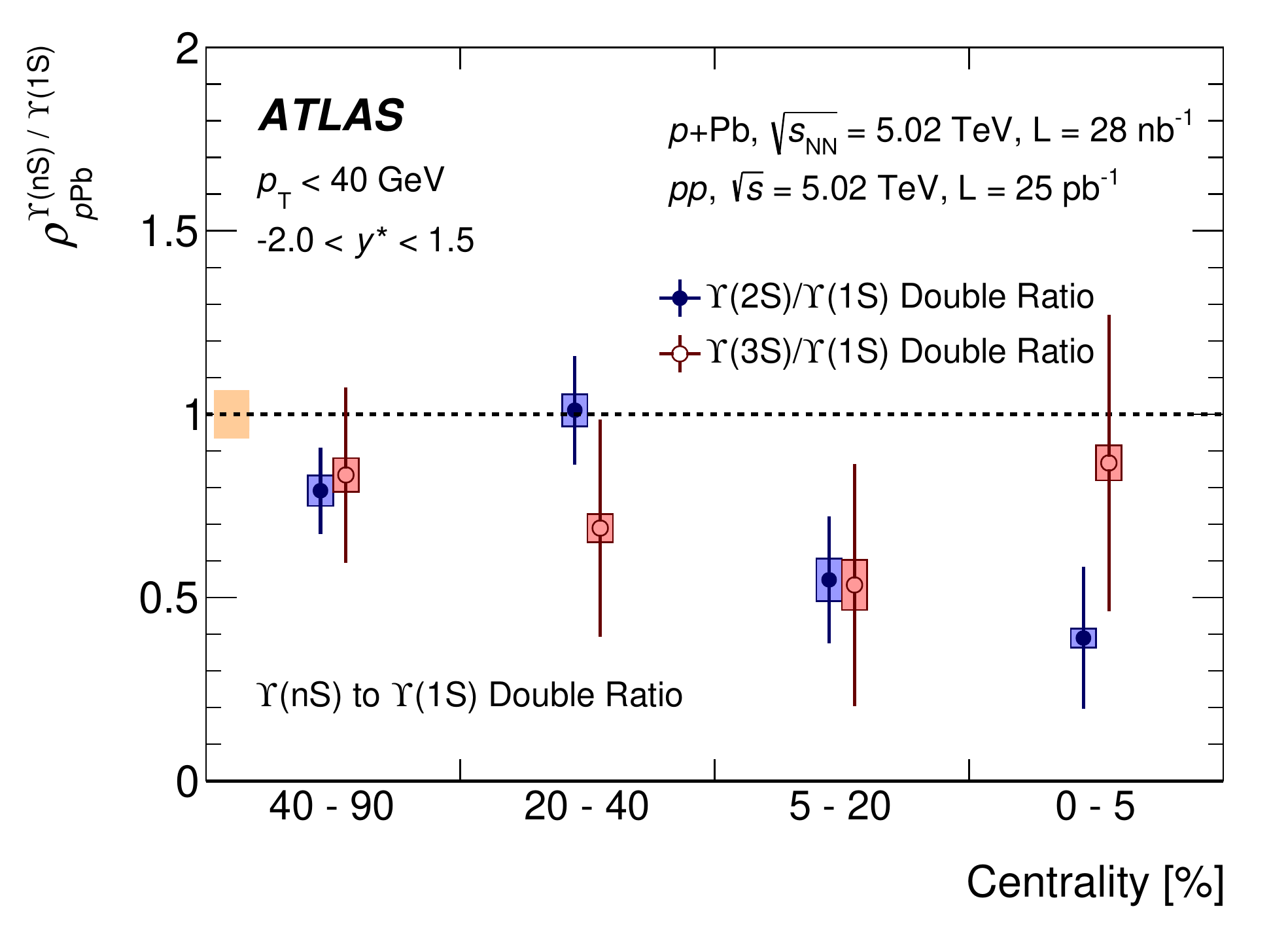}
  \caption{The prompt charmonium double ratio, $\rhopbc$, (left) and the bottomonium double ratio, $\rhopbb$, (right) as a function of event centrality in \pPb\ collisions .
  The vertical error bars correspond to the statistical uncertainties,
  and the vertical sizes of coloured boxes around the data points represent the uncorrelated systematic uncertainties in \pPb\ collisions.
  The vertical size of the leftmost yellow box around $\rhopb = 1$ represents the total uncertainty of the $pp$ reference which is the same for all centralities.
  }
  \label{fig:UpsiDR_cent}
  \end{center}
\end{figure}

The prompt $\psip$ to $\jpsi$ production double ratio, \rhopbc, is shown in Figure~\ref{fig:DR_EtRap} in intervals of $y^*$.
A decreasing trend with one-standard-deviation significance of the double ratio is observed from backward to forward centre-of-mass rapidity.
The $\pt$ and $y^*$ integrated bottomonium double ratios, \rhopbb ($n=2,3$) are shown in Figure~\ref{fig:DR_Upsi}.
Both the integrated \rhopbbii and \rhopbbiii are found to be less than unity by two standard deviations,
and they are consistent with each other within the uncertainties.
The double ratio as a function of centrality is shown in Figure~\ref{fig:UpsiDR_cent}.
Both \rhopbc and \rhopbbii are found to decrease slightly with increasing centrality 
at the significance level of one-standard-deviation, 
while conclusions about \rhopbbiii are precluded by the size of the statistical uncertainties.

\clearpage
\section{Summary}
\label{sec:conclusion}

The double-differential production cross sections of five quarkonium states,
$\Jpsi$, $\psip$, $\Upsins$ ($n = 1, 2, 3$) are measured using 
using \pPb (\pp) collision data corresponding to an integrated luminosity of $28~\inb$ ($25~\ipb$)
at a centre-of-mass energy per nucleon pair of $5.02~\TeV$ collected by the ATLAS experiment at the LHC.
The measured prompt charmonium production cross section in the range of $8 < \pt < 40~\GeV$ is found to be compatible with non-relativistic QCD predictions, 
while only the bottomonium results at $\pt > 15~\GeV$ can be described by the non-relativistic QCD predictions.
The measured non-prompt production cross sections of $\Jpsi$ and $\psip$ in \pp collisions are found to 
be consistent with fixed-order next-to-leading-logarithm calculations.

The nuclear modification factors of prompt and non-prompt $\jpsi$ in \pPb\ collisions, $\rpb$,
measured for $8 < \pt < 40~\GeV$ are found to be consistent with unity,
and no apparent dependence on $\pt$ or rapidity is observed in the measured range, 
which indicates weak modification of $\jpsi$ production due to cold nuclear matter effects at central rapidity and high $\pt$.
The $\rpb$ for $\Upsi$ is measured for $\pt < 40~\GeV$ and is found to be smaller than unity at $\pt < 15~\GeV$,
increasing with $\pt$ and becoming compatible with unity at high $\pt$.
The observed suppression of $\Upsi$ production in \pPb collisions at low $\pt$ 
suggests that the nuclear parton distribution functions are modified relative to those of the nucleon.
No apparent rapidity dependence of $\Upsi~\rpb$ is observed.
The production ratios of prompt and non-prompt $\Jpsi$ to $Z$ boson are found to be constant in bins of centrality. 
As the $Z$ boson production in \pPb collisions was found to scale with the number of binary collisions after applying centrality bias correction factors, 
the same conclusion can be drawn for $\Jpsi$ production in \pPb collisions.
The self-normalised yields of ground-state quarkonium states in \pPb collisions are found to correlate linearly with self-normalised event activity 
expected for events with the highest event activity where the self-normalised yields show two-standard-deviation departure from the linear correlation trend.

The prompt charmonium double ratio is found to decrease slightly from the backward to the forward centre-of-mass rapidity.
The prompt \psip production is suppressed with respect to prompt \jpsi production in \pPb collisions with a significance of one standard deviation.
The production of excited bottomonium states, $\Upsii$ and $\Upsiii$,
is found to be suppressed with respect to $\Upsi$ in the integrated kinematic ranges
of $\pt < 40~\GeV$ and $-2 < y^* < 1.5$
in \pPb\ collisions with significance at the level of two standard deviations.
Both the prompt $\psip$ to $\Jpsi$ and $\Upsii$ to $\Upsi$ double ratios show decreasing behaviour in more central collisions.
The decreasing trends from peripheral to central are at the significance level of one standard deviation.
A stronger cold nuclear matter effect is observed in excited quarkonium states compared to that in ground states.

This work expands the kinematic range of measured charmonium and bottomonium cross sections in $pp$ and \pPb\ collisions.
It thus serves as an additional dataset for constraining different models of cold nuclear matter effects and quantifying heavy quarkonium production.
In particular, the behaviour of the ground-state yields as a function of centrality is found to match that of $Z$ bosons, 
while excited states are relatively suppressed in more central collisions.

\section*{Acknowledgements}

We thank CERN for the very successful operation of the LHC, as well as the
support staff from our institutions without whom ATLAS could not be
operated efficiently.

We acknowledge the support of ANPCyT, Argentina; YerPhI, Armenia; ARC, Australia; BMWFW and FWF, Austria; ANAS, Azerbaijan; SSTC, Belarus; CNPq and FAPESP, Brazil; NSERC, NRC and CFI, Canada; CERN; CONICYT, Chile; CAS, MOST and NSFC, China; COLCIENCIAS, Colombia; MSMT CR, MPO CR and VSC CR, Czech Republic; DNRF and DNSRC, Denmark; IN2P3-CNRS, CEA-DRF/IRFU, France; SRNSF, Georgia; BMBF, HGF, and MPG, Germany; GSRT, Greece; RGC, Hong Kong SAR, China; ISF, I-CORE and Benoziyo Center, Israel; INFN, Italy; MEXT and JSPS, Japan; CNRST, Morocco; NWO, Netherlands; RCN, Norway; MNiSW and NCN, Poland; FCT, Portugal; MNE/IFA, Romania; MES of Russia and NRC KI, Russian Federation; JINR; MESTD, Serbia; MSSR, Slovakia; ARRS and MIZ\v{S}, Slovenia; DST/NRF, South Africa; MINECO, Spain; SRC and Wallenberg Foundation, Sweden; SERI, SNSF and Cantons of Bern and Geneva, Switzerland; MOST, Taiwan; TAEK, Turkey; STFC, United Kingdom; DOE and NSF, United States of America. In addition, individual groups and members have received support from BCKDF, the Canada Council, CANARIE, CRC, Compute Canada, FQRNT, and the Ontario Innovation Trust, Canada; EPLANET, ERC, ERDF, FP7, Horizon 2020 and Marie Sk{\l}odowska-Curie Actions, European Union; Investissements d'Avenir Labex and Idex, ANR, R{\'e}gion Auvergne and Fondation Partager le Savoir, France; DFG and AvH Foundation, Germany; Herakleitos, Thales and Aristeia programmes co-financed by EU-ESF and the Greek NSRF; BSF, GIF and Minerva, Israel; BRF, Norway; CERCA Programme Generalitat de Catalunya, Generalitat Valenciana, Spain; the Royal Society and Leverhulme Trust, United Kingdom.

The crucial computing support from all WLCG partners is acknowledged gratefully, in particular from CERN, the ATLAS Tier-1 facilities at TRIUMF (Canada), NDGF (Denmark, Norway, Sweden), CC-IN2P3 (France), KIT/GridKA (Germany), INFN-CNAF (Italy), NL-T1 (Netherlands), PIC (Spain), ASGC (Taiwan), RAL (UK) and BNL (USA), the Tier-2 facilities worldwide and large non-WLCG resource providers. Major contributors of computing resources are listed in Ref.~\cite{ATL-GEN-PUB-2016-002}.

\printbibliography
\clearpage 
\begin{flushleft}
{\Large The ATLAS Collaboration}

\bigskip

M.~Aaboud$^\textrm{\scriptsize 137d}$,
G.~Aad$^\textrm{\scriptsize 88}$,
B.~Abbott$^\textrm{\scriptsize 115}$,
J.~Abdallah$^\textrm{\scriptsize 8}$,
O.~Abdinov$^\textrm{\scriptsize 12}$$^{,*}$,
B.~Abeloos$^\textrm{\scriptsize 119}$,
S.H.~Abidi$^\textrm{\scriptsize 161}$,
O.S.~AbouZeid$^\textrm{\scriptsize 139}$,
N.L.~Abraham$^\textrm{\scriptsize 151}$,
H.~Abramowicz$^\textrm{\scriptsize 155}$
% ...
\bigskip
\\
% ...
$^{8}$ Department of Physics, The University of Texas at Arlington, Arlington TX, United States of America\\
% ...
$^{12}$ Institute of Physics, Azerbaijan Academy of Sciences, Baku, Azerbaijan\\
% ...
$^{88}$ CPPM, Aix-Marseille Universit{\'e} and CNRS/IN2P3, Marseille, France\\
% ...
$^{115}$ Homer L. Dodge Department of Physics and Astronomy, University of Oklahoma, Norman OK, United States of America\\
% ...
$^{119}$ LAL, Univ. Paris-Sud, CNRS/IN2P3, Universit{\'e} Paris-Saclay, Orsay, France\\
% ...
$^{137}$ $^{(a)}$ Facult{\'e} des Sciences Ain Chock, R{\'e}seau Universitaire de Physique des Hautes Energies - Universit{\'e} Hassan II, Casablanca; $^{(b)}$ Centre National de l'Energie des Sciences Techniques Nucleaires, Rabat; $^{(c)}$ Facult{\'e} des Sciences Semlalia, Universit{\'e} Cadi Ayyad, LPHEA-Marrakech; $^{(d)}$ Facult{\'e} des Sciences, Universit{\'e} Mohamed Premier and LPTPM, Oujda; $^{(e)}$ Facult{\'e} des sciences, Universit{\'e} Mohammed V, Rabat, Morocco\\
% ...
$^{139}$ Santa Cruz Institute for Particle Physics, University of California Santa Cruz, Santa Cruz CA, United States of America\\
% ...
$^{151}$ Department of Physics and Astronomy, University of Sussex, Brighton, United Kingdom\\
% ...
$^{155}$ Raymond and Beverly Sackler School of Physics and Astronomy, Tel Aviv University, Tel Aviv, Israel\\
% ...
$^{161}$ Department of Physics, University of Toronto, Toronto ON, Canada\\
% ...
$^{a}$ Also at Department of Physics, King's College London, London, United Kingdom\\
$^{b}$ Also at Institute of Physics, Azerbaijan Academy of Sciences, Baku, Azerbaijan\\
$^{c}$ Also at Novosibirsk State University, Novosibirsk, Russia\\
$^{d}$ Also at TRIUMF, Vancouver BC, Canada\\
$^{e}$ Also at Department of Physics {\&} Astronomy, University of Louisville, Louisville, KY, United States of America\\
% ...
$^{*}$ Deceased
\end{flushleft}


\end{document}